\documentstyle[art12]{article}

\newfont{\AMSL}{msbm10 scaled \magstep 1}

\def\RA{\mbox{\AMSL R}}

\newcommand{\ch}{\mathop{\rm ch}\nolimits}

\newtheorem{de}{Definition}[section]
\newtheorem{lem}{Lemma}[section]
\newtheorem{th}{Theorem}[section]
\newtheorem{pr}{Proposition}[section]
\newtheorem{cor}{Corollary}[section]
\newtheorem{rem}{Remark}[section]

\title{Propagation and interaction of shock waves of quasilinear equation
\thanks{This work was supported by Grants 99-01-00719, 99-01-01074,
and 99-01-39128 of Russian Foundation for Basic Research.}\/}

\author{V.~G.~Danilov\\
\small Moscow Technical University of Communication and Informatics\\
\small danilov@amath.msk.ru\\
\and
\and
V.~M.~Shelkovich\\
\small St.-Petersburg State Architecture and Civil Engineering University,
Russia\\
\small shelkv@svm.abu.spb.ru\\}

\date{}

\begin{document}
\maketitle

\begin{abstract}
We propose a new regularization method for constructing a shock
wave type solution with nonsmooth front (interaction of shock
waves)  for quasilinear equations in the one-dimensional case.
\end{abstract}

\setcounter{equation}{0}

\section{Introduction and Main Results.\/}
\label{s1}

    1. Consider the quasilinear first order equation
\begin{equation}
\label{1}
L[u]=u_t+(f(u))_x=0,
\end{equation}
where $f(u)$ is a smooth function of at most polynomial growth,
$u= u(x,t)$, \ $x \in \RA$, \ $t\in (0, \ T)$, with the initial
value $u(x,0)= u^{0}(x)$. The Hopf equation
\begin{equation}
\label{2}
L_H[u]= u_t+(u^2)_x= 0.
\end{equation}
is a special case of this equation.

   It is known that equation (\ref{1}) can have a discontinuous shock
wave type solution even in the case of the smooth initial condition
$u^{0}(x)$. Such piecewise smooth solutions are determined by integral
identities, namely, the function $u(x,t)$ is called a generalized
solution of equation (\ref{1}) in the domain $\Omega \subset \RA^2$
if for all functions with compact supports
$\varphi(x,t) \in {\cal D}(\Omega)$ we have
\begin{equation}
\label{2*}
\int_{\Omega }
\Big[u(x,t)\varphi_t(x,t)+f(u(x,t))\varphi_x(x,t)\Big]\,dx\,dt= 0.
\end{equation}

   O.~A.~Oleinik~\cite{1} proved the existence of a solution of the
Cauchy problem for equation (\ref{1}) in the class of piecewise smooth
functions provided that certain additional stability conditions hold.
A theory of such problems for quasilinear equations and systems of
equations is described in~\cite{1} --~\cite{6}

    However, the Cauchy problem can be posed for equation (\ref{1}) with
the singular initial value $u^{0}(x)$ if its singularity is stronger
than that of the Heaviside function
$$
H(\xi)= \left\{
\begin{array}{rcl}
1, \ \xi &>&0, \\
0, \ \xi &<&0,\\
\end{array}
\right.
$$
i.e. a jump.

    Thus, in~\cite{7} --~\cite{9} the problem about the propagation of
an
infinitely narrow $\delta$-soliton was considered. In the simplest case
nonlinear waves of this type arise if we consider the weak asymptotics,
as $\varepsilon  \to +0$, of the one-soliton solution
$$
u(x,t,\varepsilon )
= \frac{3v}{2}\ch^{-2}(\frac{\sqrt{v}}{2}(x-vt)/\varepsilon ),
$$
of the Korteweg-de-Vries equation (KdV)
$$
L_{KdV}[u]= u_t + (u^2)_x + \varepsilon ^2u_{xxx} =  0.
$$
Up to $O_{{\cal D}'}(\varepsilon ^2)$, this asymptotics gives the
{\it infinitely narrow $\delta$-soliton\/}
$$
u_{\varepsilon }(x,t)= A\varepsilon \delta(x-vt), \quad \varepsilon \to +0,
$$
where $\delta(x)$ is the Dirac $\delta$-function,
$A=\frac{3v}{2}\int \ch^{-2}(\xi)\,d\xi= 6\sqrt{v}$ and
by $O_{{\cal D}'}(\varepsilon ^{\alpha})$ we denote a distribution from
${\cal D}'(\RA)$ such that for any test function $\varphi(x) \in {\cal D}$
$$
\langle O_{{\cal D}'} (\varepsilon ^{\alpha}), \varphi(x)\rangle
= O(\varepsilon ^{\alpha}),
$$
and $O(\varepsilon ^{\alpha})$ is understood in the ordinary sense.
We stress that here and below we consider all distributions as
distributions depending on the argument~$x$. All other arguments
are considered as  parameters.

    Since $u(x,t,\varepsilon )= O_{{\cal D}'}(\varepsilon )$ and
$\varepsilon ^2u_{xxx}= O_{{\cal D}'}(\varepsilon ^3)$ as
$\varepsilon  \to +0$, the limit expression $u_{\varepsilon }(x,t)$
was interpreted by V.~P.~Maslov and V.~A.~Tsupin~\cite{7}, V.~P.~Maslov
and G.~A.~Omelyanov ~\cite{8} as a {\it generalized solution \/}
(asymptotic up to $O_{{\cal D}'}(\varepsilon ^2)$ of the Hopf equation
(\ref{2}) which is {\it the limit problem} for the KdV equation.
In~\cite{7},~\cite{8} the corresponding generalized Hugoniot conditions
similar to those at a shock wave front were obtained. Here the
initial value $u^{0}(x)= A\varepsilon \delta(x)$,\ $\varepsilon  \to+0$
is not a distribution but {\it an asymptotic distribution\/} (see~\cite{9}
and below). In ~\cite{9} asymptotic generalized solutions of the quasilinear
equations (\ref{1}) and systems of equations were considered in the form
of infinitely narrow $\delta$-solitons in the algebra of {\it asymptotic
distributions\/}.

   In~\cite{9} the problem on propagation of infinitely narrow
$P$-solitons was also considered. This is a new type of nonlinear waves
which arise in solving the Cauchy problem for equation (\ref{1}) with
the initial value in the form of the asymptotic distribution
$u^{0}(x)= u_0^{0}(x)+g^{0}(x)\varepsilon P(x^{-1})$, \ $\varepsilon \to +0$,
where $P(x^{-1})$ is the principal value of the function $x^{-1}$.

    In~\cite{10} --~\cite{12} problems on propagation and interaction of
$\delta$-waves for semilinear hyperbolic systems are studied, i.e.
the Cauchy problem is solved for the initial value $u^{0}(x)$ whose
singularity is of the Dirac $\delta$-function type.

    Generally speaking, the class of problems involving determination
of singular solutions of quasilinear and semilinear equations can lead
to the problem to define multiplication of distributions (generalized
functions) and to construct associative algebras containing
distributions.
In particular, if we rewrite equation (\ref{1}) as $u_t+f'(u)u_x= 0$
which is equivalent to the divergent form for the case of smooth solutions,
then even for piecewise smooth solutions of this equation with jumps
there arises a problem to define the product of the Dirac
$\delta$-function by the Heaviside function $\delta(\xi)H(\xi)$.

    These problems require a development of special analytic methods.
There exist various approaches to their solution~\cite{9},~\cite{11},
~\cite{13} --~\cite{19}.

   In~\cite{18},~\cite{19}, and especially in~\cite{9} was developed
a new analytical method, {\it the weak asymptotic method\/}, which
enables investigation of {\it the dynamics of propagation\/} of various
types of singularities of quasilinear differential equations and
hyperbolic first order systems. The fundamental ideas of this method
originate from the papers by Y.~B.~Livchak~\cite{15}, Li Bang-He~\cite{16},
V.~K.~Ivanov~\cite{17} (where the product of distributions was defined
as the weak asymptotics of the product of approximations of
distributions being multiplied with respect to the approximation parameter)
and V.~P.~Maslov~\cite{6} (where the direct substitution of a singular
ansatz into a quasilinear equation was used).

In the present paper the weak asymptotics method is applied to
the investigation of the dynamics of
both propagation and interaction of initial discontinuities, i.e. shock
waves. Further we describe the essence of the weak asymptotics
method and give the definitions~\ref{de2},~\ref{de3} of a generalized
asymptotic solution and a generalized solution to equation (\ref{1}).

    In Theorem~\ref{th2} and Corollary~\ref{cor1}, Corollary~\ref{cor2},
using the weak asymptotics method, we write out a system of equations
to determine the dynamics of propagation of {\it one\/} shock wave for
equation (\ref{1}) and the Hopf equation (\ref{2}), i.e. in the class
of piecewise smooth functions we solve the Cauchy problem with the
following initial value:
\begin{equation}
\label{3}
u^{*0}(x)= \left\{
\begin{array}{rcl}
u_0^0(x)+e^0(x), \quad  x &<& x_0, \\
u_0^0(x),        \quad  x &>& x_0, \\
\end{array}
\right.
\end{equation}
where $x_0$ is the initial position of the shock wave front, $u_0^0(x)$,
\ $e^0(x)>0$ are smooth functions. Analogous results for strictly
hyperbolic first order systems were derived in~\cite{9}.
The proof of Theorem~\ref{th2} is given in Section~\ref{s3}.

    In Theorem~\ref{th7} we write a system of equations to determine
the dynamics of propagation and interaction of {\it two\/} initial
discontinuities (shock waves with constant amplitudes) for the Hopf
equation (\ref{2}) is considered since it is a typical model of a
nonlinear hyperbolic equation. In Theorem~\ref{th8} the results
obtained are extended to the case of equation (\ref{1}). Namely, we
solve, in the sense of Definition~\ref{de3}, the Cauchy problem for
equations (\ref{2}) and (\ref{1}), in the class of piecewise constant
functions, with the following initial condition:
\begin{equation}
\label{3*}
u^{*0}(x)= \left\{
\begin{array}{rcl}
u_0+e_1+e_2, \quad && x < x_1^0, \\
u_0+e_2,     \quad && x_1^0< x <x_2^0, \\
u_0,         \quad &&  x > x_2^0, \\
\end{array}
\right.
\end{equation}
where $u_0$, \ $e_1$, \ $e_2$ are constants. The solution of the Cauchy
problem is given by the formula which can be used for all $t\ge 0$.

    To prove Theorems~\ref{th7},~\ref{th8} we construct asymptotic
solutions of equations (\ref{2}) and (\ref{1}) in the sense of
Definition~\ref{de2}. These asymptotic solutions are given by
Theorems~\ref{th3},~\ref{th4} proved in Sections~\ref{s4},~\ref{s5}.

    In Theorems~\ref{th3},~\ref{th4} we describe the dynamics of
the shock wave merging process.

    In what follows we assume that $f^{\prime\prime}(u) \ge 0$ and the
solutions $u(x,t)$ are piecewise smooth functions satisfying the Oleinik
stability conditions at every point $(x,t)$ of a discontinuity
line~\cite{1}:
\begin{equation}
\label{3**}
u(x-0,t)>u(x+0,t).
\end{equation}

    The one-phase and two-phase generalized solutions {\it constructed\/}
in Theorems~\ref{th2},~\ref{th7},~\ref{th8} in the sense of
Definition~\ref{de3} are generalized solutions in the sense of the
standard definition by the integral identity (\ref{3}). Thus, provided
that the stability conditions are fulfilled, the Oleinik uniqueness
theorem holds for them~\cite{1}.

    Note that our methods are also applicable for solving a similar
problem in the case of systems of quasilinear first order equations.

    2. {\bf The weak asymptotics method.\/} When solving the problem on
propagation and interaction of singularities of quasilinear equations one
has to extend the Schwartz distribution space. Therefore, in~\cite{9} we
constructed an associative algebra of {\it asymptotic distributions\/}
with the identity and free of zero divisors generated by the linear span
of (one-dimensional) associated homogeneous distributions. The elements
$f^*_{\varepsilon }(x)$ from the algebra of asymptotic distributions
are defined as weak asymptotic expansions of elements $f^*(x,\varepsilon )$
from the linear span of approximations of one-dimensional associated
homogeneous distributions as the approximation parameter $\varepsilon $
tends to zero. Each element of this algebra has a unique representation
in the form of the asymptotics (in the weak sense) whose coefficients
are distributions.

    The {\it product\/} of associated homogeneous distributions is
defined as a weak asymptotic expansion of the product of approximations of
multiplied distributions as $\varepsilon \to+0$. This product is an
element of the associative algebra of asymptotic distributions.

   In particular, infinitely narrow $\delta$ and $P$-solitons are
elements of the algebra of asymptotic distributions.

    Now turn to the description of our technique omitting the algebraic
aspects which are given in detail in~\cite{9} and~\cite{19}.

    To study the interaction of (two) singularities of equation (\ref{1}),
we solve the initial value problem with distribution initial data
$$
u^{*0}(x)= u_0^0(x)+\sum_{j= 1}^2e_j^0(x)s_j(x-x_j),
$$
where $u_0^0(x)$, \ $e_j^0(x)$ are smooth functions, $s_j(\xi)$ are
distributions (generalized functions) or asymptotic distributions
(see~\cite{9},~\cite{19}) and $x_j$ are constants.

    The solution of this initial value problem is found in the
form of the singular ansatz
\begin{equation}
\label{4}
u^*_{\varepsilon }(x,t)= u_0(x,t)+\sum_{j= 1}^2e_j(x,t)s_j(x-\phi_j(t))
\end{equation}
where $u_0(x,t)$, \ $e_j(x,t)$, \ $\phi_j(t)$ are functions to be
found and in the general case $\phi_j(t)$, $e_j(x,t)$ and
$s_j(\xi)$ may depend on the small parameter $\varepsilon $. The
singular ansatz $u^*_{\varepsilon }(x,t)$ belongs to the associative
and commutative differential algebra of asymptotic distributions,
However, actually, we do not use this fact in this paper directly,
since to define the associative and commutative product of
distributions, we need to calculate all terms of the asymptotic
expansion (i.e., to calculate the weak asymptotics of the product of
approximations up to $O_{{\cal D}'}(\varepsilon^\infty)$). But in view
of the above remarks and Definition~\ref{de3} below, to construct a
generalized asymptotic solution, it is sufficient to calculate the weak
asymptotics of the above-mentioned product of approximations up to
$O_{{\cal D}'}(\varepsilon^\infty)$, $N= 1$ or $N= 2$ (see below).

    To study the interaction of two nonlinear waves we assume:
1) $s_j(\xi)= \delta(\xi)$, \ $j= 1,2$ in the case of the $\delta$-waves;
2) $s_j(\xi)= H(\xi)$, \ $j= 1,2$ in the case of the shock waves;
3) $s_{j}(\xi)= {\varepsilon }\delta(\xi)$, \ $j= 1,2$ and
$s_{j}(\xi)= {\varepsilon }H(\xi)$, \ $j= 3,4$ in the case of infinitely
narrow $\delta$-solitons;
4) $s_j(\xi)= {\varepsilon }P(\xi^{-1})$, \ $j=1,2$ in the
case of infinitely narrow $P$-solitons.

    To find the solution of the form $u_{\varepsilon }(x,t)$ we construct
the smooth ansatz
\begin{equation}
\label{4*}
u^*(x,t,\varepsilon )= u_0(x,t)
+\sum_{j= 1}^2e_j(x,t)s_j\big(x-\phi_j(t, \varepsilon ), \varepsilon \big),
\end{equation}
which is a smooth approximation of the singular ansatz (\ref{4}).
Here $s_j(\xi,\varepsilon )$ is a smooth approximation of the
distribution or asymptotic distribution $s_j(\xi)$ and $\varepsilon $
is the approximation parameter.

    To construct the approximations $s_j(\xi, \varepsilon )$ we use
the fact that a correspondence can be set up between each distribution
(generalized function) $f(x)\in {\cal D}'$ and its approximation:
\begin{equation}
\label{5}
f(x,\varepsilon )= f(x)*K(x,\varepsilon )=
\langle f(t), K(x-t,\varepsilon )\rangle, \quad \varepsilon  >0,
\end{equation}
where $*$ is a convolution, the kernel
$K(x,\varepsilon )= \frac{1}{\varepsilon }
\omega\big(\frac{x}{\varepsilon }\big)$ is a $\delta$-type function such
that $0\le \omega(z)\in C^{\infty}(\RA)$, \ $\int\omega(z)\,dz= 1$, \
$\omega(z)$ has a compact support or decreases sufficiently rapidly as
$|z|\to \infty$, for example, $|\omega(z)|\le C(1+|z|)^{-N}$ for all
positive integers $N$~\cite[ch.I, \S 4.6.]{20}.

   For all test functions $\varphi(x) \in {\cal D}$ we have:
$$
\lim\limits_{\varepsilon  \to +0}\langle f(x,\varepsilon ),\varphi(x) \rangle
=\langle f(x),\varphi(x) \rangle.
$$

    For example, for an approximation of $\delta$-function we have from
(\ref{5}):
\begin{equation}
\label{5*}
\delta(x,\varepsilon )= \frac{1}{\varepsilon }\omega(\frac{x}{\varepsilon }).
\end{equation}

    For an approximation of the Heaviside function $H(x)$ we have
from (\ref{5}):
$$
H(x,\varepsilon )= H(x)*\frac{1}{\varepsilon }
\omega\big(\frac{x}{\varepsilon }\big)
= \int_{0}^{\infty}\omega\big(\frac{x}{\varepsilon }-t\big)\,dt.
$$
Hence we find
\begin{equation}
\label{5**}
H(x,\varepsilon )= \omega_0(\frac{x}{\varepsilon })
= \int_{-\infty}^{\frac{x}{\varepsilon }}\omega(\eta)\,d\eta,
\end{equation}
where $\omega_0(z)\in C^{\infty}(\RA)$, \
$\lim_{z \to +\infty}\omega_0(z)= 1$, \ $\lim_{z \to -\infty}\omega_0(z)= 0$.

    The smooth ansatz $u(x,t,\varepsilon )$ is substituted into the
equation (\ref{1}) and the weak asymptotics (in terms of the space of
the distributions ${\cal D}'$) up to $O_{{\cal D}'}(\varepsilon ^N)$,
where $N= 1$ for delta and shock waves, $N= 2$ for infinitely narrow
solitons, is found of the left-hand side of this equation as
$\varepsilon \to +0$. We define the {\em generalized solution}
({\em asymptotics solution}) to equation (\ref{1}) as the
{\em weak asymptotics} $u_{\varepsilon }(x,t)$ of the smooth ansatz
$u(x,t,\varepsilon )$, as $\varepsilon \to +0$.

    {\it The key role\/} in solving the problem of the interaction of
nonlinear waves of the equation (\ref{1}) plays the construction of the
weak asymptotics of the smooth function from the smooth ansatz (\ref{4*}),
which can be represented in the form
$$
f\big(u_0(x,t)
+\sum_{j= 1}^2e_j(x,t)s_j\big(x-\phi_j(t, \varepsilon ),\varepsilon \big)
\qquad\qquad\qquad\qquad\qquad\qquad
$$
\begin{equation}
\label{6}
=f(u_0(x,t))
+\sum_{j= 1}^2B_j\big(u_0(x,t),e_1(x,t),e_2(x,t),\rho\big)s_j(x-\phi_j(t))
+ O_{{\cal D}'}(\varepsilon ^N),
\end{equation}
where the estimate $O_{{\cal D}'}(\varepsilon ^N)$ is uniform with
respect to $\psi(t,\varepsilon )$, where
$\psi(t,\varepsilon )= \phi_2(t,\varepsilon )-\phi_1(t,\varepsilon )$,
$\rho= \frac{\psi(t,\varepsilon )}{\varepsilon }$, and {\it the interaction
switches\/} $B_j\big(u_0(x,t),e_1(x,t),e_2(x,t),\rho\big)$, $j=1,2$ which
are smooth functions, can be computed explicitly.

    Expansion (\ref{6}) allows one {\it to separate the singularities\/}
of the ansatz. Therefore, substituting expansion (\ref{6}) and the singular
ansatz (\ref{4}) into a quasilinear equation and setting equal to zero the
coefficients of the different powers of the small parameter $\varepsilon $
and of the linear independent distributions we obtain a system of equations
(in particular, the Rankine--Hugoniot type condition) which describes
the dynamics of singularities and defines the smooth functions
$u_0(x,t)$, \ $e_j(x,t)$, \ $\phi_j(t)$, \ $j= 1,2$.

    If the propagation of one singularity is studied we set $e_2^0(x)= 0$
in the initial value $u^{*0}(x)$, and we set $e_2(x,t)= 0$ in (\ref{4}),
(\ref{4*}).

    3. {\bf The weak asymptotics method and shock waves.\/} It follows
from (\ref{4}), (\ref{4*}) that to study propagation of a shock wave,
i.e. to solve the Cauchy problem (\ref{1}), (\ref{3}), we must seek a
solution in the form of the singular ansatz
\begin{equation}
\label{7}
u^*_{\varepsilon }(x,t)= u_0(x,t)+e(x,t)H(x-\phi(t))
\end{equation}
and substitute into equation (\ref{1}) the smooth ansatz
\begin{equation}
\label{9}
u^*(x,t,\varepsilon )= u_0(x,t)+e(x,t)H(-x+\phi(t),\varepsilon ),
\end{equation}
where $u_0(x,t)$, \ $e(x,t)$, \ $\phi(t)$ are functions to be
found, $\phi(0)= x_0$ and $H\big(\xi,\varepsilon \big)$ is the approximation
(\ref{5**}) of the Heaviside function $H(\xi)$.

    It should be noted that an the algebraic approach to the construction
of solutions of the shock wave type, that is, the method of direct
substitution the singular ansatz (\ref{7}) into a quasilinear equation
in divergent form was first used by V.~P.~Maslov~\cite{6}. His approach
was based on the fact that a smooth function~$f$ of the sum
$u_0(x)+e(x)H(x)$, where $u_0(x)$, \ $e(x)$ are smooth functions, can
be represented in the form as its argument:
$$
f\big(u_0(x)+e(x)H(x)\big)= f\big(u_0(x)\big)
+\Big[f\big(u_0(x)+e(x)\big)-f\big(u_0(x)\big)\Big]H(x).
$$
After substituting this expression into a quasilinear first order
equation written in the divergent form, and differentiating, one can
obtain a linear combination of the unit, the Dirac $\delta$-function
and the Heaviside function with smooth coefficients. Setting this
expression to zero and separating the singularities, we obtain the
relations which determine the solution.

    In order to investigate the interaction of two shock waves, that is,
to solve the Cauchy problem (\ref{1}), (\ref{3*}), we should seek a solution
in the form of the singular ansatz
\begin{equation}
\label{8}
u^*_{\varepsilon }(x,t)
=u_0(x,t)+\sum_{k=1}^2\Big[e_k(x,t)H\big(-x+\phi_k(t,\varepsilon )\big)\Big],
\end{equation}
and substitute the smooth ansatz
\begin{equation}
\label{10}
u^*(x,t,\varepsilon )= u_0(x,t)+\sum_{k= 1}^2
\Big[e_k(x,t)H_k\big(-x+\phi_k(t,\varepsilon ),\varepsilon \big)\Big],
\end{equation}
into equation (\ref{1}). Here $u_0$(x,t), \ $e_k(x,t)$ are smooth functions
to be found, $\lim_{\varepsilon \to+0}\phi_k(0,\varepsilon )= x_k^0$,
\ $H_k\big(\xi,\varepsilon \big)$ are approximations (\ref{5**})
of the Heaviside function $H(\xi)$, \ $k=1,2$.

\begin{de}
\label{de1} \rm
    Let $f(u)$ be a smooth function, and
let $u^*_{\varepsilon }(x,t)$ is an asymptotic distribution
(singular ansatz) (\ref{7}) or (\ref{8}), and $u^*(x,t,\varepsilon )$
is a smooth ansatz (\ref{9}) or (\ref{10}).

    By the $O_{{\cal D}'}(\varepsilon )$-substitution we call the weak
asymptotics of the expression $f\big(u^*(x,t,\varepsilon )\big)$ as
$\varepsilon  \to +0$ computed up to $O_{{\cal D}'}(\varepsilon )$. We
denote this substitution $f\big(u^*_{\varepsilon }(x,t)\big)$.
\end{de}

    Now we present the {\it main lemma\/} which gives an asymptotic
expansion of the type (\ref{6}) for the case of shock waves. The proof
of Lemma~\ref{lem2} is given in Section~\ref{s2}.

\begin{lem}
\label{lem2}
    Let
$$
H(-x,\varepsilon )
= \int_{-\infty}^{\frac{-x}{\varepsilon }}\omega_1(\eta)\,d\eta,
\quad\quad
H(-x+a,\varepsilon )
= \int_{-\infty}^{\frac{-x+a}{\varepsilon }}\omega_2(\eta)\,d\eta
$$
be approximations of the Heaviside functions $H(-x)$, \ $H(-x+a)$,
respectively, the approximating functions $\omega_k(z) \in C^{\infty}(\RA)$
are nonnegative and either have compact supports or decrease sufficiently
rapidly, as $|z|\to \infty$ and $\int\omega_k(z)\,dz=1$, \ $k=1,2$.

   Let $f(u)$ be a smooth function of at most polynomial growth and
let $u_0(x,t)$, \ $e_1(x,t)$, \ $e_2(x,t)$ be bounded functions. Then
we have the asymptotics:
$$
f\big(u_0(x,t)+e_1(x,t)H_1(-x,\varepsilon )
+e_2(x,t)H_2(-x+a,\varepsilon )\big)= f(u_0(x,t))
\qquad\qquad\qquad\qquad\qquad
$$
$$
+\Big[f(u_0(x,t)+e_1(x,t))-f(u_0(x,t))\Big]H(-x)
+\Big[f(u_0(x,t)+e_1(x,t))-f(u_0(x,t))\Big]H(-x+a)
$$
\begin{equation}
\label{65}
+B_1(x,t,\frac{a}{\varepsilon })H(-x)+B_2(x,t,-\frac{a}{\varepsilon })H(-x+a)
+O_{{\cal D}'}(\varepsilon ), \quad \varepsilon  \to +0,
\end{equation}
where the estimate $O_{{\cal D}'}(\varepsilon )$ is uniform with respect
to $a$.

    The functions $B_k(x,t,\rho)$, \ $k= 1,2$ called {\bf interaction switch
functions\/} have the following form
\begin{equation}
\label{66}
\begin{array}{rcl}
B_1(x,t,\rho)&= &\int\Big[f'\big(u_0(x,t)+e_1(x,t)\omega_{01}(-\eta)
+e_2(x,t)\omega_{02}(-\eta+\rho)\big) \\
\qquad&&\qquad\qquad\qquad
-f'\big(u_0(x,t)+e_1(x,t)\omega_{01}(-\eta)\big)\Big]
e_1(x,t)\omega_{1}(-\eta)\,d\eta, \\
B_2(x,t,-\rho)&= &\int\Big[f'\big(u_0(x,t)+e_1(x,t)\omega_{01}(-\eta-\rho)
+e_2(x,t)\omega_{02}(-\eta)\big) \\
\qquad&&\qquad\qquad\qquad
-f'\big(u_0(x,t)+e_2(x,t)\omega_{02}(-\eta)\big)\Big]
e_2(x,t)\omega_{2}(-\eta)\,d\eta. \\
\end{array}
\end{equation}

    In addition, the interaction switch functions satisfy the following
relations: for each $\rho \in \RA$
\begin{equation}
\label{67}
\begin{array}{rcl}
B_1(x,t,\rho)+B_2(x,t,-\rho)&= &f(u_0(x,t)+e_1(x,t)+e_2(x,t))
-f(u_0(x,t)+e_1(x,t)) \\
\qquad&&\quad\qquad\qquad\qquad
-f(u_0(x,t)+e_2(x,t))+f(u_0(x,t)), \\
\lim_{\rho\to+\infty}B_k(x,t,\rho)&= &f(u_0(x,t)+e_1(x,t)+e_2(x,t))
-f(u_0(x,t)+e_1(x,t)) \\
\qquad&&\quad\qquad\qquad\qquad
-f(u_0(x,t)+e_2(x,t))+f(u_0(x,t)),\\
\lim_{\rho\to-\infty}B_k(x,t,\rho)&= &0, \qquad k= 1,2. \\
\end{array}
\end{equation}
\end{lem}

\begin{rem}
\label{rem22} \rm
    For simplicity, we have selected the approximating functions
$\omega_k(z)$ such that they either have compact supports or decrease
sufficiently rapidly as $|z|\to \infty$, for example,
$|\omega_k(z)|\le C_k(1+|z|)^{-N}$ for all positive integers $N$,
\ $k= 1,2$. Therefore, we have
$$
\begin{array}{rcl}
\displaystyle
\omega_{0k}(z)&= &\int_{-\infty}^{z}\omega_k(\eta)\,d\eta= 1+O(z^{-N}),

\quad z \to +\infty,\\
\displaystyle
\omega_{0k}(z)&= &O(|z|^{-N}), \quad z \to -\infty. \\
\end{array}
$$

    Consequently, applying the Lagrange theorem we obtain the estimate
$$
[f'\big(u_0+e_1\omega_{01}(-\eta)+e_2\omega_{02}(-\eta+\rho)\big)
-f'\big(u_0+e_1\omega_{01}(-\eta)\big)\Big]e_1\omega_{1}(-\eta)
\qquad\qquad
$$
$$
= f^{\prime\prime}\big([u_0+e_1\omega_{01}(-\eta)]
+\Theta e_2\omega_{02}(-\eta+\rho)\big)
e_1e_2\omega_{1}(-\eta)\omega_{02}(-\eta+\rho), \quad 0<\Theta <1.
$$

    It follows from this estimate and (\ref{66}) that
$$
B_1(x,t,\rho)= f(u_0+e_1+e_2)-f(u_0+e_1)-f(u_0+e_2)+f(u_0)+O(\rho^{-N}),
\quad \rho \to +\infty,
$$
$$
B_1(x,t,\rho)= O(|\rho|^{-N}), \quad \rho \to -\infty,
\qquad\qquad\qquad\qquad\qquad\qquad\qquad
$$
where $N= 1,2,\dots$; if $\rho_0$ is a constant then
$$
B_1(x,t,\rho)= B_1(\rho_0)+O(\rho-\rho_0), \quad \rho \to \rho_0.
$$
\end{rem}

\begin{cor}
\label{cor21}
   Let $f(u)$ be a smooth function of at most polynomial growth and let
$u_0(x,t)$, \ $e(x,t)$ be bounded functions and $H(-x,\varepsilon )$
be approximations of the Heaviside functions $H(-x)$. Then
$$
f\big(u_0(x,t)+e(x,t)H(-x+\phi(t),\varepsilon )\big)=f\big(u_0(x,t)\big)
\qquad\qquad\qquad\qquad\qquad\qquad\qquad\qquad
$$
\begin{equation}
\label{11}
\qquad
+\Big[f\big(u_0(x,t)+e(x,t)\big)-f\big(u_0(x,t)\big)\Big]H(-x+\phi(t))
+ O_{{\cal D}'}(\varepsilon ), \quad \varepsilon  \to +0.
\end{equation}
\end{cor}

    If we set $e_1(x,t)= e(x,t)$, \ $e_2(x,t)= 0$ in (\ref{66}) then
$B_1(x,t,\rho)= B_2(x,t,-\rho)= 0$ and expansion (\ref{11}) immediately
follows from (\ref{65}).

\begin{rem}
\label{rem21} \rm
    In view of the obvious relation
\begin{equation}
\label{12}
[H(x,\varepsilon )]^r= H(x) + O_{{\cal D}'}(\varepsilon ),
\quad \varepsilon  \to +0, \qquad r= 2,3,\dots,
\end{equation}
for the case when $f(u)$ is a polynomial, formula (\ref{11}) was
proved in~\cite[8.1.]{9}.
\end{rem}

\begin{cor}
\label{cor22}
    Let
$$
H(-x,\varepsilon )
= \int_{-\infty}^{\frac{-x}{\varepsilon }}\omega_1(\eta)\,d\eta,
\quad\quad
H(-x+a,\varepsilon )
= \int_{-\infty}^{\frac{-x+a}{\varepsilon }}\omega_2(\eta)\,d\eta
$$
be the approximations of the Heaviside functions $H(-x)$, \ $H(-x+a)$,
respectively, where the functions $\omega_k(z) \in C^{\infty}(\RA)$ either
have compact supports or decrease sufficiently rapidly for $|z|\to \infty$,
and $\int\omega_k(z)\,dz= 1$, \ $k=1,2$.

    Then
\begin{equation}
\label{25}
H(-x,\varepsilon )H(-x+a,\varepsilon )
= B_1\big(\frac{a}{\varepsilon }\big)H(-x) +
B_2\big(-\frac{a}{\varepsilon }\big)H(-x+a) + O_{{\cal D}'}(\varepsilon ),
\quad \varepsilon  \to +0,
\end{equation}
where the estimate $O_{{\cal D}'}(\varepsilon )$ is uniform with respect
to $a$.

    The functions $B_k(\rho)$ called {\bf interaction switch functions\/},
have the form
\begin{equation}
\label{26}
\begin{array}{rcl}
\displaystyle
B_1(\rho) &=&
\int_{-\infty}^{\rho}\big({\check \omega}_1*\omega_2\big)(\eta)\,d\eta, \\
\displaystyle
B_2(-\rho) &=&
\int_{-\infty}^{-\rho}\big(\omega_1*{\check \omega}_2\big)(\eta)\,d\eta, \\
\end{array}
\end{equation}
where ${\check \omega}(\eta)= \omega(-\eta)$, \ $*$ is the operation of
convolution.

    In this case,
\begin{equation}
\label{27}
\begin{array}{rcl}
\displaystyle
B_1(\rho)+B_2(-\rho)&= &1, \quad \rho \in \RA, \\
\displaystyle
B_k(\infty)= \lim_{\rho \to +\infty}B_k(\rho)&= &1, \\
\displaystyle
B_k(-\infty)= \lim_{\rho \to -\infty}B_k(\rho)&= &0. \\
\end{array}
\end{equation}
\end{cor}

    To prove Corollary~\ref{cor22} we find the asymptotics of the expression
$$
H(-x,\varepsilon )H(-x+a,\varepsilon )
\qquad\qquad\qquad\qquad\qquad\qquad\qquad\qquad\qquad\qquad\qquad\qquad\
qquad
$$
$$
=\frac{1}{4}\Big(f\big(H_1(-x,\varepsilon )+H_2(-x+a,\varepsilon )\big)-
f\big(H_1(-x,\varepsilon )\big)-f\big(H_2(-x+a,\varepsilon )\big)\Big),
$$
where $f(u)= u^2$.

    Setting in (\ref{65}) $u_0(x,t)= 0$, \ $e_1(x,t)= e_2(x,t)= 1$ and
taking into account relation (\ref{12}), we obtain (\ref{25}). In this
case it follows from (\ref{66}) and (\ref{5**}):
$$
B_k\big((-1)^{k-1}\rho\big)
= \int\omega_{03-k}(-\eta+(-1)^{k-1}\rho)\big)\omega_{k}(-\eta)\,d\eta
$$
$$
= \int_{-\infty}^{\infty}\bigg[\int_{-\infty}^{-\eta+(-1)^{k-1}\rho}
\omega_{3-k}(v)\,dv\bigg]\omega_k(-\eta)\,d\eta, \quad k= 1,2.
$$

    After the substitution $v= z-\xi$ in the second integral, we reduce
$B_k(\rho)$ to the form
$$
B_k\big((-1)^{k-1}\rho\big)= \int_{-\infty}^{(-1)^{k-1}\rho}\bigg[
\int_{-\infty}^{\infty}\omega_k(-\xi)\omega_{3-k}(z-\xi)\,d\xi\bigg]\,dz
$$
and thus obtain (\ref{26}).

    In this case (\ref{67}) implies (\ref{27}).

    4. {\bf Generalized solutions.\/} Let us define a generalized
discontinuous solution and rules for substituting ansatzs of the
type (\ref{7}), (\ref{8}) into equation (\ref{1}) (see Definition~\ref{de1}).
\begin{de}
\label{de2} \rm
    Let $u^*(x,t,\varepsilon )$ be the smooth ansatz (\ref{9}), (\ref{10}).
We call the corresponding asymptotic distribution (singular ansatz)
(\ref{7}), (\ref{8}), $u^*_{\varepsilon }(x,t)$, by a {\sl generalized
asymptotic\/} shock wave type solution of the equation $L[u]= 0$,
for $t \in [0, \ T]$, with the initial condition
$u^{0*}_{\varepsilon }(x)$, if the following relations hold:
$$
\begin{array}{rcl}
\displaystyle
L[u^*(x,t,\varepsilon )] &= & O_{{\cal D}'}(\varepsilon ), \\
\displaystyle
u^*_{\varepsilon }(x,0) &= & u^{0*}_{\varepsilon }(x)
+O_{{\cal D}'}(\varepsilon ), \\
\end{array}
$$
where the first estimate is uniform with respect to $t \in [0, \ T]$.
\end{de}

\begin{de}
\label{de3} \rm
    Let $u^*_{\varepsilon }(x,t)$ be an asymptotic of the order of
$O_{{\cal D}'}(\varepsilon )$ solution of the equation $L[u]= 0$
with the initial condition $u^{0*}_{\varepsilon }(x)$. By a
{\sl generalized solution\/} of this equation with the initial
condition $u^{0*}_{\varepsilon }(x)$ we call the weak limit
$$
u^*(x,t)= \lim_{\varepsilon \to +0}u^*_{\varepsilon }(x,t).
$$
\end{de}

    After substitution of expansion (\ref{11}) and the singular ansatz
(\ref{7}) (or expansion (\ref{65}) and the singular ansatz (\ref{8}))
in equation (\ref{1}), {\it to separate the singularities\/} of the
left hand side of equation (\ref{1}) and to obtain a system of equations
which describes the dynamics of singularities and determines the smooth
functions $u_0(x,t)$, \ $e_k(x,t)$, \ $\phi_k(t)$, \ $k= 1,2$
we use the following Lemma.

\begin{lem}
\label{lem4}
    If $A(x)$, \ $B_k(x)$, \ $C_k(x)$ are smooth functions $k= 1,2$ and
$a>0$. Then
$$
A(x)+B_1(x)\theta(x)+C_1(x)\delta(x)+B_2(x)\theta(x-a)+C_2(x)\delta(x-a)=0
$$
if and only if
$$
\begin{array}{rclrcl}
A(x)&= &0,               \quad \mbox{for\/} \,    &x& <0, \\
A(x)+B_1(x)&= &0,        \quad \mbox{for\/} \, 0< &x& <a, \\
A(x)+B_1(x)+B_2(x)&= &0, \quad \mbox{for\/} \,    &x& >a, \\
C_1(0)&= &0, \\
C_2(a)&= &0. \\
\end{array}
$$
\end{lem}

    5. {\bf Propagation of shock wave of equation (\ref{1}).\/}

    Let us consider the propagation of the single shock wave of the
equation $L[u]= u_t+(f(u))_x= 0$ with the initial value (\ref{3}).
In the framework of our approach we shall seek the solution in the
form of the singular ansatz (\ref{7}) which coincides with the initial
value $u^{*0}(x)$ for $t= 0$. This means that one substitute the
smooth ansatz (\ref{9}) approximating (\ref{7}) into the equation $L[u]=0$
and find the weak asymptotics, up to $O {{\cal D}'}(\varepsilon)$, of the
left hand side of the equation, as $\varepsilon  \to+0$.

\begin{th}
\label{th2}
     Consider the Cauchy problem {\rm (\ref{1})}, {\rm (\ref{3})},
where $f(u)$ is a smooth function,
$f^{\prime\prime}(u)\ge 0$ and $u_0^0(x)$, \ $e^0(x)>0$ are smooth functions. Suppose
there exists $T>0$, such that
\begin{equation}
\label{19}
\begin{array}{rcl}
\displaystyle
-\inf_{\xi>\phi(0)}[u^{0\prime}(\xi)f^{\prime\prime}(u^0(\xi))] &<& T^{-1},\\
\displaystyle
-\inf_{\xi<\phi(0)}\Big[u^0(\xi)+e^0(\xi)\Big]'
f^{\prime\prime}(u^0(\xi)+e^0(\xi))] &<& T^{-1}. \\
\end{array}
\end{equation}

    Then in the sense of Definitions~{\rm \ref{de2}},~{\rm \ref{de3}}
equation {\rm (\ref{1})} for $t \in [0, \, T]$ has a discontinuous
solution of the form {\rm (\ref{7})}
$$
u^*(x,t)=u_0(x,t) + e(x,t)H(-x+\phi(t)),
$$
if and only if the unknown functions $u_0(x,t)$, \ $e(x,t)$, \
$\phi(t)\in C^{\infty}$ satisfy the following system of equations
\begin{equation}
\label{20}
\begin{array}{rcl}
\displaystyle
L[u_0(x,t)] &= &0, \qquad x>\phi(t), \\
\displaystyle
L[u_0(x,t)+e(x,t)] &= &0, \qquad x<\phi(t), \\
\displaystyle
\frac{d\phi(t)}{dt}&= &
\frac{f\big(u_0(x,t)+e(x,t)\big)-f\big(u_0(x,t)\big)}{e(x,t)}
\bigg|_{x= \phi(t)},\\
\end{array}
\end{equation}
where $u^*(x,0)= [u_0(x,t)+e(x,t)]\Big|_{t= 0}= u_0^0(x)+e^0(x)$ for
$x < \phi(0)$ and $u^*(x,0)= u_0(x,t)\Big|_{t= 0}= u_0^0(x)$ for
$x > \phi(0)$, $\phi(0)= x_0$.
\end{th}

\begin{cor}
\label{cor1}
   In the case of the Hopf equation $L_H[u]= u_t+(u^2)_x= 0$ system
{\rm (\ref{20})} has the form
\begin{equation}
\label{21}
\begin{array}{rcl}
\displaystyle
L_H[u_0(x,t)]&=&0, \qquad x>\phi(t), \\
\displaystyle
L_H[u_0(x,t)+e(x,t)]&=&0, \qquad x<\phi(t), \\
\displaystyle
\frac{d\phi(t)}{dt}&=&2u_0(\phi(t),t)+e(\phi(t),t). \\
\end{array}
\end{equation}
\end{cor}

    The last equations in systems (\ref{20}) and (\ref{21}) represent
the Rankine-Hugoniot conditions along the discontinuity line
$x=\varphi(t)$ for shock waves~\cite[Ch. 4, \S 1.2.]{5}:
\begin{equation}
\label{22}
\frac{d x}{d t}=\frac{[f(u)]}{[u]},
\end{equation}
where
$$
\begin{array}{rcl}
\displaystyle
[u] &=& u(\varphi(t)+0,t) - u(\varphi(t)-0,t), \\
\displaystyle
[f(u)] &=& f(u(\varphi(t)+0,t)) - f(u(\varphi(t)-0,t)) \\
\end{array}
$$
are jumps of functions $u(x,t)$ and $f(u)$, respectively, along the
discontinuity line $x= \varphi(t)$. In our case, $u= u_0+e$ to the left
of the discontinuity, and $u=u_0$ to the right of the discontinuity.

\begin{cor}
\label{cor2}
    Solution {\rm (\ref{7})} constructed in the sense of
Definition~{\rm \ref{de2}} to the Cauchy problem for equation
{\rm (\ref{1})} with the initial value {\rm (\ref{3})}, that is,
a function of the form $u^*(x,t)= u_0(x,t)+ e(x,t)H(-x+\phi(t))$,
whose smooth components satisfy system of equations {\rm (\ref{20})}
from Theorem~{\rm \ref{th2}}, is the unique generalized solution of
the same Cauchy problem in the sense of the standard definition
by integral identity {\rm (\ref{2*})}.
\end{cor}

    {\it Proof.} To prove that the solution of the Cauchy problem in the
sense of Definition~\ref{de3} is the solution in the sense of the
standard definition by the integral identity (\ref{2*}) we remind that in
view of Theorem~\ref{th2}, the approximation $u^*(x,t,\varepsilon )$
(\ref{9}) of the solution (\ref{7}) satisfies the equation
$L[u^*(x,t,\varepsilon )]= O_{{\cal D}'}(\varepsilon )$.
Let us apply the left-hand and right-hand sides of this relation to
an arbitrary test function $\varphi(x,t)\in{\cal D}(\Omega)$, \
$\Omega\subset R^2$.

    Since for $\varepsilon >0$ the functions $u^*(x,t,\varepsilon )$
is smooth, integrating by parts, we obtain
$$
\int_{\Omega}\Big[u^*(x,t,\varepsilon )\varphi_t(x,t)
+f\big(u^*(x,t,\varepsilon )\big)\varphi_x(x,t)\Big]\,dxdt
=\langle O_{{\cal D}'}(\varepsilon ),\varphi(x,t)\rangle=O(\varepsilon).
$$
Since the limit $u^*(x,t)$ of the family $u^*(x,t,\varepsilon )$ as
$\varepsilon \to +0$ in the ${\cal D}'$ sense is a bounded locally
integrable function, we have $\int_{\Omega}\Big[u^*(x,t)\varphi_t(x,t)
+f\big(u^*(x,t)\big)\varphi_x(x,t)\,dxdt= 0$, as $\varepsilon \to+0$,
which coincides with the usual integral identity (\ref{2*})
(see~\cite{1}--~\cite{5}).

    Due to our assumptions, the stability condition (\ref{3**}) is
satisfied and by Theorem of O.~A.~Oleinik this solution is unique~\cite{1}.

    6. {\bf Interaction of the shock waves of the Hopf equation (\ref{2}).\/}

    To study the interaction of two shock waves of the Hopf equation
(\ref{2}) $L_H[u]= u_t+(u^2)_x= 0$ we shall seek its solution in the
form of the singular ansatz (\ref{8}) which, for $t= 0$ coincides with
distribution (\ref{3*}).

    First we construct an asymptotic solution of problem (\ref{2}),
(\ref{3*}) in the sense of Definition~\ref{de2}, which is given by
Theorem~\ref{th3} proved in Sections~\ref{s4}.

\begin{th}
\label{th3}
    For all $t\in [0, \ +\infty)$ in the sense of Definition~{\rm \ref{de2}}
there exists the asymptotic solution
$$
u^*(x,t,\varepsilon )= u_0+
\sum_{k= 1}^2e_kH_k\big(-x+\phi_k(t,\varepsilon ),\varepsilon \big),
$$
for the Hopf equation {\rm (\ref{2})} such that:

    {\rm 1)} for all $t > 0$: $L_H[u^*(x,t,\varepsilon )]
= O_{{\cal D}'}(\varepsilon )$,

    {\rm 2)} $u^{*0}(x)-u^*(x,0,\varepsilon )= O_{{\cal D}'}(\varepsilon )$,
where
$$
u^{*0}(x)= \left\{
\begin{array}{rcl}
u_0+e_1+e_2, \quad && x < x_1^0, \\
u_0+e_2,     \quad && x_1^0 < x < x_2^0, \\
u_0,         \quad && x > x_2^0, \\
\end{array}
\right.
$$
$x_1^0 < x_2^0$ and $u_0$, \ $e_1>0$, \ $e_2>0$ are constants;

    {\rm 3)} for $t\in (0, \ t^*)$
$$
\lim_{\varepsilon \to +0}u^*(x,t,\varepsilon )
=u_0+\sum_{k=1}^2e_kH\big(-x+\phi_{k0}(t)\big),
$$
where
$$
\begin{array}{rcl}
\displaystyle
\phi_{10}(t)=\lim_{\varepsilon \to +0}\phi_1(t,\varepsilon )
&=&\phi_{10}(0)+(2u_0 + e_1 + 2e_2)t, \\
\displaystyle
\phi_{20}(t)=\lim_{\varepsilon \to +0}\phi_2(t,\varepsilon )
&= &\phi_{20}(0)+(2u_0 + e_2)t, \\
\end{array}
$$
$\phi_{k0}(0)=x_k^0$, \ $k= 1,2$, $t^*= \frac{x_2^0-x_1^0}{e_1+e_2}$
is the shock wave merging time;

    {\rm 4)} for $t\in (t^*, \ \infty)$
$$
\lim_{\varepsilon \to +0}u^*(x,t,\varepsilon )
= u_0+(e_1+e_2)H\big(-x+\widehat{\phi}_{-}(t)\big),
$$
where
$$
\widehat{\phi}_{-}(t)=\lim_{\varepsilon \to +0}\phi_k(t,\varepsilon )
= x^* + (2u_0 + e_1 + e_2)(t-t^*), \quad k= 1,2,
$$
$$
x^*=\phi_{k0}(t^*)=\frac{(2u_0 + e_1 + 2e_2)x_2^0-(2u_0 + e_2)x_1^0}
{e_1 + e_2}.
$$

    Here
$$
H_k(-x,\varepsilon )
= \int_{-\infty}^{\frac{-x}{\varepsilon }}\omega_k(\eta)\,d\eta,
$$
are approximations of the Heaviside function $H(-x)$, where
$\omega_k(z)$ are smooth functions which either have compact supports or
decrease sufficiently rapidly as $|z|\to \infty$, \ $\int\omega_k(z)\,dz=1$,
\ $k=1,2$.

    The phases $\phi_k(t,\varepsilon )$ have the form
$$
\phi_k(t,\varepsilon )= \phi_{k0}(t)
+\psi_0(t)\phi_{k1}(\tau)\Big|_{\tau=\frac{\psi_{0}(t)}{\varepsilon }},
$$
where $\psi_0(t)= \phi_{20}(t)-\phi_{10}(t)$ and the perturbations of
phase functions are given by formulae {\rm (\ref{51})}:
$$
\phi_{k1}(\tau)=(-1)^{k-1}\frac{2e_{3-k}}{(e_1+e_2)\tau}\int_{0}^{\tau}
\Big(1-B_1\big(\rho(\tau')\big)\Big)\,d\tau'.
$$
According to {\rm (\ref{26})},
$$
B_1(\rho)
=\int_{-\infty}^{\rho}\big({\check\omega}_1*\omega_2\big)(\eta)\,d\eta,
$$
and $\rho=\rho(\tau)
=[\phi_2(t,\varepsilon)-\phi_1(t,\varepsilon)]/\varepsilon$ is a
solution of the differential equation with the
boundary condition {\rm (\ref{44})}, {\rm (\ref{45})}:
$$
\begin{array}{rcl}
\displaystyle
\frac{d\rho}{d\tau} &= & 2B_1(\rho)-1, \\
\displaystyle
\frac{\rho(\tau)}{\tau}\Big|_{\tau \to +\infty}&= &1.\\
\end{array}
$$
\end{th}

    Note that, in fact, Theorem~\ref{th3} is based on the remark that
the variable~$t$ is considered as a parameter. If we considered the
variable~$t$ as a variable having the same rights as the variable~$x$,
we also had to calculate a weak asymptotics with respect to the
variable~$t$. In this case we would obtain Heaviside functions instead
of interaction switch functions and could not obtain a formula that were
uniform in~$t$ for the solution $u^*(x,t,\varepsilon)$.

    Theorem~\ref{th3} implies the following result.

\begin{th}
\label{th7}
    The Cauchy problem for the Hopf equation {\rm (\ref{2})},
{\rm (\ref{3*})} in the sense of Definition~{\rm \ref{de3}} has a
discontinuous solution of the form {\rm (\ref{8})}
\begin{equation}
\label{53}
u^*(x,t)= u_0+\sum_{k= 1}^2\Big[e_kH\big(-x+\phi_k(t)\big)\Big],
\end{equation}
where
\begin{equation}
\label{54}
\begin{array}{rcl}
\displaystyle
\phi_{1}(t)&= &\phi_{10}(0)+(2u_0+e_1+2e_2)t-e_2(t-t^*)H(t-t^*), \\
\displaystyle
\phi_{2}(t)&= &\phi_{20}(0)+(2u_0+e_2)t+e_1(t-t^*)H(t-t^*) \\
\end{array}
\end{equation}
are phases.

    This solution of the initial value problem {\rm (\ref{2})},
{\rm (\ref{3*})} is the unique generalized solution of this Cauchy
problem in the sense of the integral identity {\rm (\ref{2*})}.
\end{th}

    {\it Proof.} It follows from Theorem~\ref{th3} that distribution
(\ref{53}) is a generalized solution of the initial value problem
(\ref{2}), (\ref{3*}) in the sense of Definition~\ref{de3}.

    In view of Theorem~\ref{th3}, the approximation $u^*_{\varepsilon}(x,t)$
of the solution (\ref{53}) uniformly in~$t$ satisfies the equation
$L[u^*(x,t,\varepsilon )]= O_{{\cal D}'}(\varepsilon )$. By applying the
left- and right-hand sides of the last relation to an arbitrary function
$\varphi(x,t)\in{\cal D}(\Omega)$, we obtain
$$
\int_{\Omega}\Big[u^*(x,t,\varepsilon )\varphi_t(x,t)
+f\big(u^*(x,t,\varepsilon )\big)\varphi_x(x,t)\Big]\,dxdt= O(\varepsilon).
$$
By integrating by parts for $\varepsilon >0$ and then passing to
the limit as $\varepsilon \to+0$, just as in the proof of Theorem~\ref{th2},
we prove that the functions $u^*(x,t)$ determined by (\ref{53})
satisfies the integral identity
$$
\int_{\Omega}\Big[u^*(x,t)\varphi_t(x,t)
+f\big(u^*(x,t)\big)\varphi_x(x,t)\Big]\,dxdt=0.
$$
Since for the piecewise constant solution $u^*(x,t)$ the stability
condition $e_k>0$, \ $k= 1,2$, holds, this solution is unique by the
Theorem of O.~A.~Oleinik~\cite{1}. Theorem~\ref{th7} is proved.

    The solution described in Theorem~\ref{th7} is of rather simple
natural form. In fact, it is sewed together of the following parts:
two shock waves till the time instant~$t^*$ and one shock wave for $t>t^*$.
We cannot say anything about the equation at the point $x=x^*$, $t= t^*$,
but we see that the trajectories of discontinuities at this point are
continuous. Moreover, one can easily see that the function (\ref{54})
itself is continuous w.r.t. to~$t$ in the weak sense.
Thus we have the following ``naive'' situation: everywhere outside
the curves  $x= \phi_k(t)$, \ $k= 1,2$, the solution (\ref{54}) can be
found as a function satisfying the equation $L[u]= 0$, while the
curves $x= \phi_k(t)$, \ $k= 1,2$, and the solution (\ref{54}) are
determined by the Rankine--Hugoniot conditions for $t\ne t^*$ and
the condition that they are continuous at the point $t= t^*$.

    However, as we shall see now, one can construct a great deal of
such points. As an example, we consider the following Cauchy problem
for the Hopf equation:
$$
u_t+(u^2)_x= 0,\quad u(x,t)\big|_{t=0}= H(-x-1)-H(x-1).
$$
One can easily verify that for $t<t^*=1$ the solution $u(x,t)$ consists
of two shock wave approaching each other:
$$
u(x,t)= H(-x+t-1)-H(x+t-1).
$$
By using the techniques used in Theorem~\ref{th3} for calculating the
solution $u^*(x,t)$, we see that the solution satisfying the integral
identity for $t\geq t^*=1$ (this solution is one of ``naive'' solutions)
is the stationary solution of the Hopf equation
$$
u(x,t)= H(-x)-H(x)
$$
consisting of two shock waves sticking together.

    Another different ``naive'' solution can be constructed as follows.
Let us consider the solution of the Hopf equation with the initial condition
$$
W(x,t)\Big|_{t=0}= \left\{
\begin{array}{ll}
H(-x-1), &\quad x\leq -1,\\
ax,&\quad -1\leq x\leq 1,\\
-H(x-1), &\quad x>1.
\end{array}
\right.
$$
The solution $W(x,t)$ of this problem has the form
$$
W(x,t)=
H(-x-\phi(t))-H(x-\phi(t))+\frac{ax}{1+2at}[1-H(-x-\phi(t))-H(x-\phi(t))].
$$
Its plot consists of two horizontal lines ($y= 1$ to the left of the point
$x_-= -\phi$ and $y= 1$ to the right of the point $x_+= \phi$) and the
inclined line $y= ax/(1+2at)$ between the points $x_-$ and $x_+$.

    At the points $x_{\pm}$ the solution has a jump whose coordinates
$x_{\pm}= \pm\phi(t)$ can be calculated from the Rankine--Hugoniot conditions
$$
\phi_t= \frac{(-1)^2-(ax/(1+2at))^2}{-1-ax/(1+2at)}\bigg|_{x=\phi}
= \frac{a\phi}{1+2at}-1,\qquad \phi(0)= 1.
$$

    Assume that $a>1$. Then, obviously, $\phi_t>0$ on some interval
$[0,t_0]$, \ $t_0>0$.

    We consider the limit of the function $W(x,t)$ in ${\cal D}'(\RA'_x)$
as $t\to-1/(2a)+0$. For any test function $\varphi(x)$ we have
$$
\lim_{t\to-1/(2a)+0}\langle W(x,t),\varphi\rangle
= \lim_{t\to-1/(2a)+0}\Big[\frac{1}{1+2at}\int^{\phi}_{-\phi}x\varphi(x)\,dx
+\int^{-\infty}_{-\phi}\varphi(x)\,dx-\int^{\phi}_{\infty}\varphi(x)\,dx\Big].
$$
By using the L'Hospital rule, we obtain
$$
\lim_{t\to-1/(2a)+0}\langle W, \varphi(x)\rangle =
\lim_{t\to-1/(2a)+0}
\Big[\frac{\phi_t(\phi\varphi(\phi))-\phi_t(\phi\varphi(-\phi))}{2a}\Big]
+\int^{-\infty}_{0}\varphi(x)\,dx-\int^{0}_{\infty}\varphi(x)\,dx
$$
$$
=\int^{-\infty}_{\infty}\big(H(-x)-H(x)\big)\varphi(x)\,dx.
$$
Consider the function
$$
U(x,t)= \left\{
\begin{array}{ll}
u(x,t)            &\quad  0\leq t\leq t^*,\\
W(x,t-1/(2a)-t^*) &\quad  t> t^*.
\end{array}
\right.
$$
This function is continuous in ${\cal D}'$ and  satisfies the classical
Hopf equation everywhere except for the discontinuity trajectories
$$
x_{\pm}= \left\{
\begin{array}{ll}
\pm(t-1)                 &\quad  t\leq t^*= 1,\\
\pm\phi(x,t-1/(2a)-t^*)  &\quad  t>1,
\end{array}
\right.
$$
which are continuous curves. Thus this is the second ``naive'' solution of
the problem of discontinuity merging. In this solution, the discontinuities
seem to interact elastically and then to run away after the interaction.

    Thus Theorem~\ref{th7} is not useless. This theorem allows one to
choose a ``true naive'' solution among all possible solutions.

    The example of solution~$W(x,t)$ is of independent interest, since
this is one of few solutions with nonmonotone motion of its singularity
\footnote{G.~A.~Omel'yanov helped the authors to clear up this situation.}.
Namely, if $a>1$, then, as was already pointed out, the discontinuities
run away from each other but, starting from the time instant at which the
inclined line $y= ax/(1+2at)$ enters the interior of the strip $y= \pm1$,
the directions of their motion are changed to the opposite, and the
discontinuities merge into the stationary solution
$$
u(x,t)= H(-x)-H(x).
$$
Moreover, at the time instant of merging, we again can ``paste in'' a
$W(x,t)$ type solution at the discontinuity point, repeat this process,
and obtain the picture of $t$-periodic elastic interaction of shock waves.

    7. {\bf Interaction of shock waves of equation (\ref{1}).\/}

    Study the process of interaction of two shock waves of the equation
$u_t+(f(u))_x= 0$, where $f(u)$ is a convex (downwards, i.e.,
$f^{\prime\prime}(u) > 0$ on the range of the solution~$u$) smooth function.
To this end, consider the Cauchy problem with the initial value (\ref{3*}).
As in the case of the Hopf equation, substitute the smooth ansatz
(\ref{10}), where $\phi_1(0)= x_1^0$, \  $\phi_2(0)= x_2^0$ and
$e_1, \ e_2>0$, approximating the singular ansatz (\ref{8}), into equation
(\ref{1}) and find the weak asymptotics of its left hand side.

    Eventually, we construct the asymptotic solution of problem (\ref{1}),
(\ref{3*}) in the sense of Definition~\ref{de2}, which is given by
Theorem~\ref{th4} proved in Sections~\ref{s5}:

\begin{th}
\label{th4}
    For all $t\in [0, \ +\infty)$ in the sense of the
Definition~{\rm \ref{de2}} there exists the asymptotic solution
$$
u^*(x,t,\varepsilon )= u_0+
\sum_{k= 1}^2e_kH_k\big(-x+\phi_k(t,\varepsilon ),\varepsilon \big),
$$
of the equation {\rm (\ref{1})} such that:

    {\rm 1)} for all $t > 0$: $L[u^*(x,t,\varepsilon )]
= O_{{\cal D}'}(\varepsilon )$,

    {\rm 2)} $u^{*0}(x)-u^*(x,0,\varepsilon )=O_{{\cal D}'}(\varepsilon )$,
where
$$
u^{*0}(x)=\left\{
\begin{array}{rcl}
u_0+e_1+e_2, \quad && x < x_1^0, \\
u_0+e_2,     \quad && x_1^0 < x < x_2^0, \\
u_0,         \quad && x > x_2^0, \\
\end{array}
\right.
$$
$x_1^0 < x_2^0$ and $u_0$, \ $e_1>0$, \ $e_2>0$ are constants;

    {\rm 3)} for $t\in (0, \ t^*)$
$$
\lim_{\varepsilon \to +0}u^*(x,t,\varepsilon )
= u_0+\sum_{k= 1}^2e_kH\big(-x+\phi_{k0}(t)\big),
$$
where
$$
\begin{array}{rcl}
\displaystyle
\phi_{10}(t)=\lim_{\varepsilon \to +0}\phi_1(t,\varepsilon )
&=&\phi_{10}(0)+\frac{f(u_0+e_1+e_2)-f(u_0+e_2)}{e_1}t, \\
\displaystyle
\phi_{20}(t)=\lim_{\varepsilon \to +0}\phi_2(t,\varepsilon )
&=&\phi_{20}(0)+\frac{f(u_0+e_2)-f(u_0)}{e_2}t, \\
\end{array}
$$
$\phi_{k0}(0)= x_k^0$, \ $k= 1,2$,
$$
t^*=e_1e_2\frac{x_2^0-x_1^0}{e_2f(u_0+e_1+e_2)-(e_1+e_2)f(u_0+e_2)+e_1f(u_0)}
$$
is the shock wave merging time;

    {\rm 4)} for $t\in (t^*, \ \infty)$
$$
\lim_{\varepsilon \to +0}u^*(x,t,\varepsilon )
= u_0+(e_1+e_2)H\big(-x+\widehat{\phi}_{-}(t)\big),
$$
where
$$
\widehat{\phi}_{-}(t)= \lim_{\varepsilon \to +0}\phi_k(t,\varepsilon )
= x^* + \frac{f\big(u_0+e_1+e_2\big)-f\big(u_0\big)}{e_1+e_2}(t-t^*),
\quad k= 1,2,
$$
$$
x^*= \phi_{k0}(t^*)= \frac{\Big[f(u_0+e_1+e_2)-f(u_0+e_2)\Big]e_2x_2^0
-\Big[f(u_0+e_2)-f(u_0)\Big]e_1x_1^0}
{e_2f(u_0+e_1+e_2)-(e_1+e_2)f(u_0+e_2)+e_1f(u_0)}.
$$

    Here
$$
H_k(-x,\varepsilon )
= \int_{-\infty}^{\frac{-x}{\varepsilon }}\omega_k(\eta)\,d\eta,
$$
are approximations of the Heaviside function $H(-x)$, where
$\omega_k(z)$ are smooth functions which either have compact supports or
decrease sufficiently rapidly as $|z|\to \infty$, \ $\int\omega_k(z)\,dz=1$,
\ $k=1,2$.

    The phases $\phi_k(t,\varepsilon )$ are set in the form
$$
\phi_k(t,\varepsilon )= \phi_{k0}(t)
+\psi_0(t)\phi_{k1}(\tau)\Big|_{\tau= \frac{\psi_{0}(t)}{\varepsilon }},
$$
where $\psi_0(t)= \phi_{20}(t)-\phi_{10}(t)$ and the perturbations of
phase functions are given by the formulas
$$
\phi_{k1}(\tau)
\qquad\qquad\qquad\qquad\qquad\qquad\qquad\qquad\qquad\qquad\qquad\qquad
\qquad\qquad\qquad\qquad\qquad
$$
$$
= (-1)^{k-1}\frac{e_{3-k}}{\tau}
\int\limits_{0}^{\tau}
\frac{f\big(u_0+e_1+e_2\big)-f\big(u_0+e_1\big)-f\big(u_0+e_2\big)
+f\big(u_0\big)-B_1(\rho(\tau'))}
{\Big[f\big(u_0+e_1+e_2\big)-f\big(u_0+e_2\big)\Big]e_2
-\Big[f\big(u_0+e_2\big)-f\big(u_0\big)\Big]e_1}\,d\tau',
$$
$$
B_1(\rho)
= \int\Big[f'\big(u_0+e_1\omega_{01}(-\eta)+e_2\omega_{02}(-\eta+\rho)\big)
-f'\big(u_0+e_1\omega_{01}(-\eta)\big)\Big]e_1\omega_{1}(-\eta)\,d\eta,
$$
where $\rho= \rho(\tau)$ has the same meaning as in Theorem~{\rm\ref{th3}}
and satisfies the differential equation with the boundary condition{\rm:}
$$
\begin{array}{rcl}
\displaystyle
\frac{d\rho}{d\tau}&= &
\frac{\Big[f(u_0+e_1+e_2)-f(u_0+e_1)\Big]e_1-\Big[f(u_0+e_1)-f(u_0)\Big]e_2
-(e_1+e_2)B_1(\rho)}
{\Big[f(u_0+e_2)-f(u_0)\Big]e_1-\Big[f(u_0+e_1+e_2)-f(u_0+e_2)\Big]e_2}, \\
\displaystyle
\frac{\rho(\tau)}{\tau}\Big|_{\tau \to +\infty}&= &1. \\
\end{array}
$$
\end{th}

    This implies the following result.

\begin{th}
\label{th8}
    The Cauchy problem for the equation {\rm (\ref{1})}, {\rm (\ref{3*})}
in the sense of Definition~{\rm \ref{de3}} has a discontinuous solution
of the form
\begin{equation}
\label{95}
u^*(x,t)= u_0+\sum_{k= 1}^2\Big[e_kH\big(-x+\phi_k(t)\big)\Big],
\end{equation}
where
\begin{equation}
\label{96}
\begin{array}{rcl}
\displaystyle
\phi_{1}(t)= \phi_{10}(0)+\frac{f(u_0+e_1+e_2)-f(u_0+e_2)}{e_1}t
\qquad\qquad\qquad\qquad\qquad\qquad\qquad\qquad \\
\displaystyle
-\bigg[\frac{f(u_0+e_1+e_2)-f(u_0+e_2)}{e_1}-
\frac{f(u_0+e_2)-f(u_0)}{e_2}\bigg]\frac{e_2}{e_1+e_2}(t-t^*)H(t-t^*),\\
\displaystyle
\phi_{2}(t)= \phi_{20}(0)+\frac{f(u_0+e_2)-f(u_0)}{e_2}t
\qquad\qquad\qquad\qquad\qquad\qquad\qquad\qquad\qquad\qquad \\
\displaystyle
+\bigg[\frac{f(u_0+e_1+e_2)-f(u_0+e_2)}{e_1}-
\frac{f(u_0+e_2)-f(u_0)}{e_2}\bigg]\frac{e_1}{e_1+e_2}(t-t^*)H(t-t^*) \\
\end{array}
\end{equation}
are phases.

    This solution of the initial value problem {\rm (\ref{1})},
{\rm (\ref{3*})} is the unique generalized solution of the Cauchy problem
in the sense of the integral identity {\rm (\ref{2*})}.
\end{th}

    This Theorem is proved in the same way as Corollary~\ref{cor2}
from Theorem~\ref{th2} and Theorem~\ref{th7}.

    8. {\bf Description of the shock wave merging process.\/}

    Thus, for $t\in (0, \ t^*)$ two shock waves whose fronts move
from the initial positions $x_1^0= \phi_1(0)$, \ $x_2^0= \phi_2(0)$, where
$x_1^0<x_2^0$ along the straight lines
$$
\begin{array}{rcl}
\displaystyle
x_1= \phi_{10}(t)&= &x_1^0+\frac{f(u_0+e_1+e_2)-f(u_0+e_2)}{e_1}t, \\
\displaystyle
x_2= \phi_{20}(t)&= &x_2^0+\frac{f(u_0+e_2)-f(u_0)}{e_2}t, \\
\end{array}
$$
with the velocities
$$
\begin{array}{rcl}
\displaystyle
\frac{d\phi_{10}(t)}{d t}&= &\frac{f(u_0+e_1+e_2)-f(u_0+e_2)}{e_1}, \\
\displaystyle
\frac{d\phi_{20}(t)}{d t}&= &\frac{f(u_0+e_2)-f(u_0)}{e_2}, \\
\end{array}
$$
where $\frac{d\phi_{10}(t)}{d t}>\frac{d\phi_{20}(t)}{d t}$.

    At the instant $t= t^*$ after the interaction shock waves merge
constituting one new shock wave which for $t\in (t^*, \ \infty)$
propagates from the point $(x^*, \ t^*)$ along the straight line
$$
x=  x^* + \frac{f\big(u_0+e_1+e_2\big)-f\big(u_0\big)}{e_1+e_2}(t-t^*)
$$
at the velocity
$$
\frac{d}{dt}\widehat{\phi}_{-}(t)
=\frac{f\big(u_0+e_1+e_2\big)-f\big(u_0\big)}{e_1+e_2},
$$
where
$$
\begin{array}{rcl}
t^* &=&e_1e_2\frac{x_2^0-x_1^0}
{e_2f(u_0+e_1+e_2)-(e_1+e_2)f(u_0+e_2)+e_1f(u_0)}, \\
x^* &=&\phi_{k0}(t^*)= \frac{\Big[f(u_0+e_1+e_2)-f(u_0+e_2)\Big]e_2x_2^0
-\Big[f(u_0+e_2)-f(u_0)\Big]e_1x_1^0}
{e_2f(u_0+e_1+e_2)-(e_1+e_2)f(u_0+e_2)+e_1f(u_0)}. \\
\end{array}
$$

\begin{rem}
\label{rem3} \rm
    The same result for the Hopf equation can be obtained using the solution
of the Burgers equation $L_B[u]= u_t+(u^2)_x+{\varepsilon }u_{xx}= 0$,
which is a regularization of the Hopf equation. It turns out that a
simple solution of the Burgers equation can be found whose limit, as
$\varepsilon \to +0$, describes the process of merging two shock waves
(see~\cite[4.7.]{6}).
\end{rem}

\section{Proof of Lemma~1.1.\/}
\label{s2}

    First, consider the case when $f(u)$ is an analytical function.
We find the weak asymptotics of the product
\begin{equation}
\label{68}
g(x,a,\varepsilon )= \omega_{01}^{m}(\frac{-x}{\varepsilon })
\omega_{02}^{n}(\frac{-x+a}{\varepsilon }),
\end{equation}
which, according to Section~\ref{s1}, approximates the product of the
distributions $H^{m}(-x)H^{n}(-x+a)$, where $m,n= 1,2,\dots$.

    It follows from the definition of the primitive of a distribution
~\cite[Ch.I, \S 2.2.]{20} that:
$$
J(a,\varepsilon )
= \Big\langle g(x,a,\varepsilon ), \varphi(x)\Big\rangle
= -\Big\langle g_{x}(x,a,\varepsilon ), \varphi^{(-1)}(x) \Big\rangle,
= J_1(a,\varepsilon )+J_2(a,\varepsilon ),
$$
where $\varphi^{(-1)}(x)= \int_{-\infty}^{x}\varphi(\xi)\,d\xi$ is the
primitive of the function $\varphi(x)$, \ $\varphi(x)\in {\cal D}'$, \
$\int_{-\infty}^{+\infty}\varphi(\xi)\,d\xi= 0$ and
$$
J_k(a,\varepsilon )
= -\Big\langle g_{kx}(x,a,\varepsilon ), \varphi^{(-1)}(x) \Big\rangle,
\qquad k=1,2,
$$
$$
g_{1x}(x,a,\varepsilon )= -m\omega_{01}^{m-1}(\frac{-x}{\varepsilon })
\omega_{02}^{n}(\frac{-x+a}{\varepsilon })
\frac{1}{\varepsilon }\omega_{1}(\frac{-x}{\varepsilon }),
$$
$$
g_{2x}(x,a,\varepsilon )= -n\omega_{01}^{m}(\frac{-x}{\varepsilon })
\frac{1}{\varepsilon }\omega_{2}(\frac{-x+a}{\varepsilon }).
$$

    Making the change of variables $x= {\varepsilon }\eta$ we have
$$
J_1(a,\varepsilon )= m\int\omega_{01}^{m-1}(-\eta)
\omega_{02}^{n}(-\eta+\frac{a}{\varepsilon })\omega_{1}(-\eta)
\varphi^{(-1)}({\varepsilon }\eta)\,d\eta.
$$
Since the integrand decreases rapidly as $|\eta| \to \infty$, we have
\begin{equation}
\label{69}
J_1(a,\varepsilon )= \varphi^{(-1)}(0)m\int\omega_{01}^{m-1}(-\eta)
\omega_{02}^{n}(-\eta+\frac{a}{\varepsilon })\omega_{1}(-\eta)\,d\eta
+ O(\varepsilon ), \quad \varepsilon  \to +0.
\end{equation}

    In an analogous way, changing variables $x= a+{\varepsilon }\eta$
we derive the asymptotics
$$
J_2(a,\varepsilon )
= \Big\langle g_{2}(x,a,\varepsilon ), \varphi(x)\Big\rangle
\qquad\qquad\qquad\qquad\qquad\qquad\qquad\qquad\qquad\qquad\qquad\qquad
\qquad
$$
\begin{equation}
\label{70}
= \varphi^{(-1)}(a)n\int\omega_{01}^{m}(-\eta-\frac{a}{\varepsilon })
\omega_{02}^{n-1}(-\eta)\omega_{2}(-\eta)\,d\eta
+ O(\varepsilon ), \quad \varepsilon  \to +0.
\end{equation}

    Summing asymptotics (\ref{69}) and (\ref{70}) we find the asymptotics
$$
J(a,\varepsilon )
=\Big\langle g(x,a,\varepsilon ), \varphi(x)\Big\rangle
\qquad\qquad\qquad\qquad\qquad\qquad\qquad\qquad\qquad\qquad\qquad\qquad
\qquad
$$
\begin{equation}
\label{71}
= B_1^{m,n}(\frac{a}{\varepsilon })\varphi^{(-1)}(0)
+B_2^{m,n}(-\frac{a}{\varepsilon })\varphi^{(-1)}(a)
+ O(\varepsilon ), \quad \varepsilon  \to +0,
\end{equation}
where
\begin{equation}
\label{72}
\begin{array}{rcl}
\displaystyle
B_1^{m,n}(\rho) &= & m\int\omega_{01}^{m-1}(-\eta)
\omega_{02}^{n}(-\eta+\rho)\omega_{1}(-\eta)\,d\eta, \\
\displaystyle
B_2^{m,n}(-\rho) &= & n\int\omega_{01}^{m}(-\eta-\rho)
\omega_{02}^{n-1}(-\eta)\omega_{2}(-\eta)\,d\eta. \\
\end{array}
\end{equation}

    Here, as in the proof of Lemma~\ref{lem2}, asymptotics (\ref{71})
can be extended to the whole space ${\cal D}$.

    Considering that
$$
\varphi^{(-1)}(0)= \int_{-\infty}^{0}\varphi(\xi)\,d\xi=
\Big\langle H(-x), \varphi(x)\Big\rangle,
$$
formula (\ref{71}) can be rewritten in the weak sense as
asymptotics (\ref{68}):
\begin{equation}
\label{73}
g(x,a,\varepsilon )= B_1^{m,n}(\frac{a}{\varepsilon })H(-x)
+B_2^{m,n}(-\frac{a}{\varepsilon })H(-x+a)
+ O_{{\cal D}'}(\varepsilon ), \quad \varepsilon  \to +0.
\end{equation}

     Using (\ref{68}), (\ref{73}), (\ref{11}) and the binomial
formula we find the weak asymptotics of the expression
$$
\Big[e_1(x,t)\omega_{01}(\frac{-x}{\varepsilon }) +
e_2(x,t)\omega_{02}(\frac{-x+a}{\varepsilon })\Big]^{n}
= e_1^n(x,t)H(-x) + e_2^n(x,t)H(-x+a)
$$
\begin{equation}
\label{74}
+{\tilde B}_1(x,t,\frac{a}{\varepsilon })H(-x)
+{\tilde B}_2(x,t,-\frac{a}{\varepsilon })H(-x+a)
+ O_{{\cal D}'}(\varepsilon ), \quad \varepsilon  \to +0.
\end{equation}

    Here, as it follows from (\ref{72}),
$$
{\tilde B}_1(x,t,\rho)
=\sum_{k= 1}^{n-1}C_{n}^{k}e_1^{k}(x,t)e_2^{n-k}(x,t)B_1^{k,n-k}(\rho)
\qquad\qquad\qquad\qquad\qquad
$$
$$
= \int\sum_{k= 1}^{n-1}nC_{n-1}^{k-1}
e_1^{k}(x,t)e_2^{n-k}(x,t)\omega_{01}^{k-1}(-\eta)
\omega_{02}^{n-k}(-\eta+\rho)\omega_{1}(-\eta)\,d\eta,
$$
where $C_{n}^{k}$ are binomial coefficients.

    The last expression can be rewritten in the form
$$
{\tilde B}_1(x,t,\rho)=
\qquad\qquad\qquad\qquad\qquad\qquad\qquad\qquad\qquad\qquad\qquad\qquad
\qquad\qquad\qquad\qquad
$$
$$
n\int\bigg[
\Big(e_1(x,t)\omega_{01}(-\eta)+e_2(x,t)\omega_{02}(-\eta+\rho)\Big)^{n-1}
-\Big(e_1(x,t)\omega_{01}(\eta)\Big)^{n-1}\bigg]
e_1(x,t)\omega_{1}(-\eta)\,d\eta.
$$

    Analogously, we obtain
$$
{\tilde B}_2(x,t,-\rho)=
\qquad\qquad\qquad\qquad\qquad\qquad\qquad\qquad\qquad\qquad\qquad\qquad
\qquad\qquad\qquad\qquad
$$
$$
n\int\bigg[
\Big(e_1(x,t)\omega_{01}(-\eta-\rho)+e_2(x,t)\omega_{02}(-\eta)\Big)^{n-1}
-\Big(e_2(x,t)\omega_{02}(-\eta)\Big)^{n-1}\bigg]
e_2(x,t)\omega_{2}(-\eta)\,d\eta.
$$

    Expanding the function $f(u)$ into the Taylor series and using
asymptotics (\ref{74}) and the formulae for ${\tilde B}_1(x,t,\rho)$, \
${\tilde B}_2(x,t,-\rho)$, we find the weak asymptotics as
$\varepsilon  \to +0$ of
$f\big(u_0+e_1H_1(-x,\varepsilon )+e_2H_2(-x+a,\varepsilon ))\big)$,
that is, obtain expansion (\ref{65}) and formulae for the interaction
switch functions (\ref{66}).

    To prove the first formula from (\ref{67}) we convert $B_1(x,t,\rho)$
from (\ref{66}) to the form
$$
B_1(x,t,\rho)= \int
f'\big(u_0+e_1\omega_{01}(-\eta)+e_2\omega_{02}(-\eta+\rho)\big)
\Big[e_1\omega_{1}(-\eta)+e_2\omega_{2}(-\eta+\rho)\Big]\,d\eta
$$
$$
-\int\Big[f'\big(u_0+e_1\omega_{01}(-\eta)+e_2\omega_{02}(-\eta+\rho)\big)
-f'\big(u_0+e_2\omega_{02}(-\eta+\rho)\big)\Big]
e_2\omega_{2}(-\eta+\rho)\,d\eta
$$
$$
-\int f'\big(u_0+e_1\omega_{01}(-\eta)\big)e_1\omega_{1}(-\eta)\,d\eta
-\int f'\big(u_0+e_2\omega_{02}(-\eta+\rho)\big)
e_2\omega_{2}(-\eta+\rho)\,d\eta.
$$

   Note that after the substitution $-\eta+\rho \to -\eta'$ the second
integral converses to the function $B_2(x,t,-\rho)$ from (\ref{66}).
Considering that $\lim_{\eta \to +\infty}\omega_{0k}(\eta)= 1$, \
$\lim_{\eta \to -\infty}\omega_{0k}(\eta)= 0$ we find that the first
integral is equal to $f\big(u_0+e_1+e_2\big)-f\big(u_0\big)$, while the
third and the forth integrals are equal, respectively, to
$f\big(u_0\big)-f\big(u_0+e_1\big)$ and $f\big(u_0\big)-f\big(u_0+e_2\big)$.

    The first formula from (\ref{67}) is thus proved.

    Since $\lim_{\rho \to +\infty}\omega_{02}(-\eta+\rho)= 1$, relation
(\ref{66}) implies:
$$
B_1(x,t,\infty)=\lim_{\rho \to +\infty}B_1(x,t,\rho)=
\int\Big[f'\big(u_0+e_1\omega_{01}(-\eta)+e_2\big)
-f'\big(u_0+e_1\omega_{01}(-\eta)\big)\Big]e_1\omega_{1}(-\eta)\,d\eta.
$$
Integrating the last expression and taking into account that
$\lim_{\eta \to +\infty}\omega_{01}(\eta)= 1$, \
$\lim_{\eta \to -\infty}\omega_{01}(\eta)= 0$, we obtain

$$
B_1(x,t,\infty)= \lim_{\rho \to +\infty}B_1(x,t,\rho)=
f\big(u_0+e_1+e_2\big)-f\big(u_0+e_1\big)-f\big(u_0+e_2\big)+f\big(u_0\big),
$$
which is the second relation from (\ref{67}). The other relations (\ref{67})
are proved in an analogous way.

    The proof of Lemma is completed by using the Theorem on
approximating a smooth function on a finite interval
by analytic ones.

\section{Proof of Theorem~1.1.\/}
\label{s3}

    Substituting $u^*_{\varepsilon }(x,t)$ into the left hand side
of equation (\ref{1}) and using (\ref{11}) from Corollary~\ref{cor21},
we have
$$
u^*_{\varepsilon t} + [f(u^*_{\varepsilon })]_x
= u_{0t}+[f(u_0)]_x + e_tH(-x+\phi) + \phi_te\delta(-x+\phi)
\qquad\qquad\qquad\qquad\qquad
$$
$$
+\Big[f(u_0(x,t)+e(x,t)\big)- f\big(u_0(x,t)\big)\Big]_xH(-x+\phi(t))
\qquad\qquad\qquad\qquad\qquad
$$
$$
-\Big[f(u_0(x,t)+e(x,t)\big)- f\big(u_0(x,t)\big)\Big]\delta(-x+\phi(t))
+ O_{{\cal D}'}(\varepsilon ), \quad \varepsilon  \to +0.
$$
Thus,
$$
L[u^*_{\varepsilon }(x,t)]= L[u_0(x,t)]+\Big\{L[u_0(x,t)+e(x,t)]
-L[u_0(x,t)]\Big\}H(-x+\phi(t))
\qquad\qquad\qquad\qquad
$$
\begin{equation}
\label{18}
+\Big\{\phi_te(x,t)-\Big[f(u_0(x,t)+e(x,t)\big)-f\big(u_0(x,t)\big)\Big]\Big\}
\delta(-x+\phi(t))+O_{{\cal D}'}(\varepsilon ), \quad \varepsilon  \to +0.
\end{equation}

    Setting the right hand side of (\ref{18}) equal to zero, using
Lemma~\ref{lem4} (to separate the singularities) and
Definitions~\ref{de2},~\ref{de3} of generalized solution, we derive
Theorem~\ref{th2} whose analog for strictly hyperbolic systems
(when $f(u)$ is a polynomial) was obtained in~\cite[8.1.]{9}.

\section{Proof of Theorem~1.2.\/}
\label{s4}

    1. In the framework of our approach, to substitute the singular
ansatz (\ref{8}) into the Hopf equation means
to substitute  the smooth ansatz (\ref{10})
$u^*(x,t,\varepsilon)$ into the Hopf equation, and then
to find a weak asymptotics, up to $O_{{\cal D}'}(\varepsilon )$,
of the left hand side of the equation $L_H[u^*(x,t,\varepsilon )]= 0$.

    According to (\ref{12}),
\begin{equation}
\label{24}
[H(-x,\varepsilon )]^2 = H(-x) + O_{{\cal D}'}(\varepsilon ),
\quad \varepsilon  \to +0.
\end{equation}

    To substitute ansatz (\ref{10}) into the equation we use asymptotics
(\ref{25}) from Corollary~\ref{cor22} for the pair product
$H(-x,\varepsilon )H(-x+a,\varepsilon )$ as $\varepsilon  \to +0$.

    In view of the aforesaid and (\ref{25}) one should involve the
dependence on $\varepsilon $ into shock wave phases, that is, seek
them in the form $\phi_k=\phi_k(t,\varepsilon )$, \ $k=1,2$.

    We consider the case when  the shock waves amplitudes $e_k(x,t)$ and
the background $u_0(x,t)$ are constant. Using (\ref{24}) -- (\ref{26}) we
find
$$
[u^*(x,t,\varepsilon )]^2 = u_0^2 + \Big(2u_0e_1+e_1^2
+2e_1e_2B_1\big(\frac{\psi(t,\varepsilon )}{\varepsilon }\big)\Big)
H\big(-x+\phi_1(t,\varepsilon )\big)
$$
$$
+ \Big(2u_0e_2+e_2^2
+ 2e_1e_2B_2\big(-\frac{\psi(t,\varepsilon )}{\varepsilon }\big)\Big)
H\big(-x+\phi_2(t,\varepsilon )\big)
+ O_{{\cal D}'}(\varepsilon ), \quad \varepsilon  \to +0,
$$
where $\psi(t,\varepsilon )= \phi_2(t,\varepsilon )-\phi_1(t,\varepsilon )$
is the distance between shock wave fronts (discontinuities).

    Substituting the singular ansatz $u^*_{\varepsilon }(x,t)$ and
the asymptotics $[u^*_{{\varepsilon }}(x,t)]^2$ into the Hopf equation
(\ref{2}) we have
$$
L_H[u^*(x,t,\varepsilon )]= \sum_{k= 1}^2
\bigg[e_k\frac{d}{dt}\phi_{k}(t,\varepsilon )
-2u_0e_k-e_k^2-2e_1e_2B_k\big((-1)^{k-1}\rho\big)\bigg]
\delta(-x+\phi_k(t,\varepsilon ))
$$
\begin{equation}
\label{30}
\qquad\qquad\qquad\qquad\qquad\qquad\qquad\qquad\qquad\qquad
+ O_{{\cal D}'}(\varepsilon ), \quad \varepsilon  \to +0,
\end{equation}
where $\rho= \frac{\psi(t,\varepsilon )}{\varepsilon }$ is the
independent variable of the interaction switch function and the
estimate $O_{{\cal D}'}(\varepsilon )$ in this representation is
uniform with respect to the distance $\psi(t,\varepsilon )$.

     From (\ref{30}), using Lemma~\ref{lem4} (to separate the singularities),
we derive the necessary and sufficient conditions for the relation
$L_H[u^*(x,t,\varepsilon )]= O_{{\cal D}'}(\varepsilon )$ to be valid:
\begin{equation}
\label{32}
\frac{d}{d t}\phi_{k}(t,\varepsilon )
= 2u_0+e_k+2e_{3-k}B_k\big((-1)^{k-1}\rho\big), \quad k= 1,2.
\end{equation}

    Before the interaction of shock waves, when $\psi(t,\varepsilon )
=\phi_2(t,\varepsilon )-\phi_1(t,\varepsilon ) > c{\varepsilon }^{1-\alpha}$,
where $c>0$, \ $0< \alpha \le 1$, according to Remark~\ref{rem22} and
(\ref{27}) we have, at least up to $O(\varepsilon )$, \
$B_1\big(\frac{\psi}{\varepsilon }\big)= 1$ and
$B_2\big(-\frac{\psi}{\varepsilon }\big)= 0$. Therefore, in this case
system (\ref{32}) becomes the system of equations which, according to
Corollary~\ref{cor1} from Theorem~\ref{th2}, describes the dynamics of
two non-interacting shock waves (taking into account that both the
amplitudes $e_k$ and the background $u_0$ are constant):
\begin{equation}
\label{36}
\begin{array}{rcl}
\displaystyle
\frac{d\phi_{10t}(t)}{d t} &= & 2u_0 + e_1 + 2e_2, \\
\displaystyle
\frac{d\phi_{20t}(t)}{d t} &= & 2u_0 + e_2, \\
\end{array}
\end{equation}
where $\phi_{10}(t)$, \ $\phi_{20}(t)$ denote the phases of non-interacting
shock waves.

    Let $\psi_0(t)= \phi_{20}(t)-\phi_{10}(t)$ be the distance between
the fronts of non-interacting shock waves. Define the interaction time
$t= t^*$ as the solution of the equation $\psi_0(t^*)= 0$.

    Equations (\ref{36}) describe the propagation of two shock waves
before the interaction time, for $t\in [0, \ t^*)$ and represent the
Hugoniot conditions (\ref{21}) along the discontinuity lines. Indeed, in
our case for the lagging shock wave we have $[u]= e_1$,
\ $[f(u)]= 2u_0e_1 + e_1^2 + 2e_1e_2$, and for the advanced one we have
$[u]= e_2$, \ $[f(u)]= 2u_0e_2 + e_2^2$.

     Thus, before the interaction, i.e. for $t\in [0, \ t^*)$, the shock
waves move with velocities (\ref{36}) along the straight lines
\begin{equation}
\label{37}
\begin{array}{rcl}
\displaystyle
x_1= \phi_{10}(t) &= & \phi_{10}(0) + (2u_0 + e_1 + 2e_2)t , \\
\displaystyle
x_2= \phi_{20}(t) &= & \phi_{20}(0) + (2u_0 + e_2)t , \\
\end{array}
\end{equation}
which intersect at the point with the following coordinates:
$$
\begin{array}{rcl}
\displaystyle
t^* &= & \frac{\phi_{20}(0)-\phi_{10}(0)}{e_1 + e_2}, \\
\displaystyle
x^* &= & \phi_{10}(0) + (2u_0 + e_1 + 2e_2)t^* , \\
\end{array}
$$
where $\phi_{k0}(0)$ are the initial positions of the discontinuities.

    2. To describe the interaction dynamics we shall seek the phases of
shock waves as functions of the fast variable
$\tau=\frac{\psi_0(t)}{\varepsilon } \in \RA$ and the slow one $t\ge 0$:
\begin{equation}
\label{38}
\phi_k(t,\varepsilon )\stackrel{def}{=}\widehat{\phi}_{k}(\tau,t)
=\phi_{k0}(t)+\psi_0(t)\phi_{k1}(\tau),
\end{equation}
where the functions $\phi_{k0}(t)$ defined by equations (\ref{37}),
for $t\in [0, \ t^*)$, {\it are extended by these equations\/} (\ref{37})
for all $t\in [t^*, \ +\infty)$. If $\tau >0$ then $t<t^*$, i.e. the
interaction has not occurred yet; if $\tau <0$ then $t>t^*$, i.e.
the interaction has occurred.

    Since we consider piecewise constant solutions, we can take the
velocities of all the shock waves formed before and after the
interaction to be constant, and seek the perturbations for the phases
$\phi_{k0}(t)$ as functions $\phi_{k1}(\tau)$ only depending on $\tau$.

    We set the following boundary values:
\begin{equation}
\label{39}
\begin{array}{rcl}
\displaystyle
\phi_{k1}(\tau)\Big|_{\tau \to +\infty} &= & 0, \\
\displaystyle
\frac{d\phi_{k1}(\tau)}{d\tau}\Big|_{\tau \to -\infty} &= & o(\tau^{-1}),\\
\end{array}
\end{equation}
That is, the derivatives of the phases with respect to the fast variable
$\tau$ tend to zero as $|\tau| \to \infty$, while the phases themselves
tend to zero as $\tau \to \infty$.

   Finding the limit values of the perturbations as $\tau \to -\infty$
$$
\phi_{k1}(\tau)\Big|_{\tau \to -\infty} = \phi_{k1,-},
$$
we find the phase limit values
$$
\widehat{\phi}_{k}(\tau,t)\Big|_{\tau \to -\infty}
= \widehat{\phi}_{k,-}(t)= \phi_{k0}(t)+\psi_0(t)\phi_{k1,-}
$$
and thus define "the result" of the interaction of shock waves for $t>t^*$.

    Let $\psi_{1}(\tau)= \phi_{21}(\tau)-\phi_{11}(\tau)$,
then the full phase difference is
$$
\psi(t,\varepsilon )= \psi_{0}(t)\big(1 + \psi_{1}(\tau)\big),
$$
the independent variable of the interaction switch function (\ref{26}) has
the form:
\begin{equation}
\label{40}
\rho(\tau)
= \frac{\psi(t,\varepsilon )}{\varepsilon }= \tau\big(1+\psi_{1}(\tau)\big).
\end{equation}

    The phase derivatives with respect to time are represented by the
following relations:
\begin{equation}
\label{41}
\frac{d}{d t}\phi_{k}(t,\varepsilon )
\stackrel{def}{=}\frac{d}{d t}\widehat{\phi}_{k}(\tau,t)
= \frac{d\phi_{k0}(t)}{d t}
+\frac{d\psi_{0}(t)}{d t}\frac{d}{d\tau}[\tau\phi_{k1}(\tau)].
\end{equation}

    Taking into account the boundary conditions (\ref{39}), we obtain
the limit values of the phases and their derivatives with respect to time
as $\tau \to -\infty$:
\begin{equation}
\label{42}
\begin{array}{rcl}
\displaystyle
\widehat{\phi}_{k,-}(t)&= &\phi_{k0}(t)+\psi_0(t)\phi_{k1,-}, \\
\displaystyle
\Big[\frac{d}{d t}\widehat{\phi}_{k}(\tau,t)\Big]_{-}
&= &\frac{d}{dt}\widehat{\phi}_{k,-}(t). \\
\end{array}
\end{equation}

    Substituting (\ref{38}), (\ref{36}) into (\ref{32}), we derive, for
all $t\ge 0$ and $\tau \in (-\infty, \ \infty)$, the following system of
equations with the boundary conditions (\ref{39}):
\begin{equation}
\label{43}
\begin{array}{rcl}
\displaystyle
\frac{d\phi_{10}(t)}{d t}
+\frac{d\psi_{0}(t)}{d t}\frac{d}{d\tau}\Big[\tau\phi_{11}(\tau)\Big]&=&
2u_0+e_1+2e_2B_1(\rho),\\
\displaystyle
\frac{d\phi_{20}(t)}{d t}
+\frac{d\psi_{0}(t)}{d t}\frac{d}{d\tau}\Big[\tau\phi_{21}(\tau)\Big]&=&
2u_0+e_2+2e_1B_2(-\rho),\\
\end{array}
\end{equation}
where $\phi_{k0}(t)$ are the functions defined, for all $t\ge, 0$ by
equations (\ref{37}).

    Subtracting one of the equations (\ref{43}) from the other we find the
differential equation satisfied by the function $\rho(\tau)$:
\begin{equation}
\label{44}
\begin{array}{rcl}
\displaystyle
\rho_{\tau}&= &F(\rho), \\
\displaystyle
\frac{\rho}{\tau}\Big|_{\tau \to +\infty}&= &1,\\
\end{array}
\end{equation}
where the boundary condition (\ref{44}) follows from (\ref{39}).

    Taking into account the first relation (\ref{27}) and the fact that,
according to (\ref{36}), $\psi_{0t}(t)= -\Big(e_1+e_2\Big)$, we find the
right hand side of the differential equation
\begin{equation}
\label{45}
F(\rho)=\frac{1}{\psi_{0t}(t)}
\Big[e_2\Big(1-2B_1(\rho)\Big)-e_1\Big(1-2B_2(-\rho)\Big)\Big]=2B_1(\rho)-1.
\end{equation}

    According to Corollary~\ref{cor22}, the function $B_1(\rho)$ is
smooth and its values belong to the interval $(0, \ 1)$. Therefore,
independent of the choice of the approximating functions $\omega_k(\xi)$,
\ $k= 1,2$, the equation $F(\rho)= 0$ has at least one root $\rho_0$
which solves the equation $B_1(\rho)= 1/2$.

    The differential equation (\ref{44}), (\ref{45}) is autonomous.
For this equation the following statement holds.
\begin{pr}
\label{pr1}
    For the autonomous equation
\begin{equation}
\label{44*}
\frac{d\rho}{d\tau}= F(\rho)
\end{equation}
with the smooth right hand side $F(\rho)$ to have a solution such that
\begin{equation}
\label{45*}
\begin{array}{rcl}
\frac{\rho(\tau)}{\tau}\Big|_{\tau \to +\infty}&= &1,\\
\displaystyle
\rho(\tau)\Big|_{\tau \to -\infty}&= &\rho_0, \\
\end{array}
\end{equation}
where $\rho_0$ is a constant, it is necessary and sufficient that the
following conditions hold:
\begin{equation}
\label{46*}
\begin{array}{rcl}
\displaystyle
F(\rho)\Big|_{\rho \to +\infty}&= &1,\\
\displaystyle
F(\rho_0)&= &0, \\
\displaystyle
F(\rho)&>&0 \,\, \mbox{for\/} \, \rho>\rho_0, \\
\end{array}
\end{equation}
where $\rho_0$ is the maximal root of the equation $F(\rho)= 0$.

    In addition, if $\rho_0$ is an ordinary (nonmultiple) root of the
equation $F(\rho)= 0$ then $\rho-\rho_0= O(\tau^{-N})$, \ $\tau \to -\infty$
for all $N= 1,2,\dots$.
\end{pr}

    {\it Proof.} Suppose that the limit relations (\ref{45*}) hold, and
${\tilde\tau}, {\tilde\tau}+1\in (\rho_0, \ +\infty)$.
Then, integrating (\ref{44*}), we have
$$
\rho({\tilde\tau}+1)-\rho({\tilde\tau})
= \int_{0}^{1}F(\rho(\tau+{\tilde\tau}))\,d\tau.
$$
Since according to the second relation (\ref{45}), the left hand side
of this equality tends to zero as ${\tilde\tau} \to -\infty$, the limit
of its right hand side is also equal to zero:
$$
\lim_{{\tilde\tau} \to -\infty}\int_{0}^{1}F(\rho(\tau+{\tilde\tau}))\,d\tau
=\int_{0}^{1}F(\rho_0)\,d\tau=F(\rho_0)= 0.
$$
That is, $\rho_0$ is a root of the equation $F(\rho_0)=0$.

    It follows from the first relation (\ref{45}) that $F(\rho)>0$ for
$\rho>\rho_0$, and consequently, $\rho_0$ is the maximal root.

    Since for $\tau_1, \tau \in (\rho_0, \ +\infty)$
\begin{equation}
\label{47*}
\frac{\rho(\tau)}{\tau}= \frac{\rho(\tau_1)}{\tau}
+\frac{1}{\tau}\int_{\tau_1}^{\tau}F(\rho(\tau'))\,d\tau',
\end{equation}
passing in (\ref{47*}) to the limit as $\tau \to +\infty$, taking into
account the first relation (\ref{45*}) and using the L'Hospital rule, we
prove that $\lim_{\rho \to +\infty}F(\rho)= 1$.

    Conversely, if (\ref{46*}) holds then $\rho= \rho_0$ is a solution of
equation (\ref{44*}). Integrating (\ref{44*}), we have
$$
\tau= \tau_1 + \int_{\rho_1}^{\rho}\frac{d\rho'}{F(\rho')}.
$$
Since the function $F(\rho)$ is smooth, the function $1/F(\rho)$ has a
non-integrable singularity at the point $\rho= \rho_0$, and the integral
$\int_{\rho_1}^{\rho}\frac{d\rho'}{F(\rho')}$ diverges as $\rho \to \rho_0$.

    That is, as $\rho$ approaches $\rho_0$, we have $\tau \to -\infty$
and therefore, $\rho= \rho_0$ is an asymptote for all integral curves within
the region $\rho_0 <\rho<+\infty$.

    The first relation (\ref{45*}) follows from (\ref{47*}).
    If $\rho_0$ is a simple root of the equation $F(\rho)= 0$ then
$F(\rho)= (\rho-\rho_0)g(\rho)\big)$, where $g(\rho)>0$ for $\rho>\rho_0$.

    The second statement is proved by passing from the differential
equation (\ref{44*}) to the differential inequality
$\frac{d\rho}{d\tau} \le (\rho-\rho_0)M$, where
$M= \max_{\rho\ge\rho_0}g(\rho)$. Integrating this inequality, we find
that $\rho-\rho_0 \le Ke^{\tau}$. Thus, $\rho \to \rho_0$ more rapidly than
any power of $|\tau|^{-1}$ as $\tau \to -\infty$.

    The proposition is proved.

    By Proposition~\ref{pr1}, as $\tau \to -\infty$, we have
$\rho= \tau\big(1+\psi_{1}(\tau)\big) \to \rho_0$, here
$B_1(\rho) \to B_1(\rho_0)= 1/2$ and $B_2(-\rho) \to B_2(-\rho_0)= 1/2$.

    Thus, passing to the limit in (\ref{43}), as $\tau \to -\infty$, and
taking into account that, according to (\ref{39}), the phase derivatives
with respect to the fast variable $\tau$ tend to zero, we derive the
following limit system of equations which describes the  shock wave
evolution after the interaction, i.e. for $t>t^*$:
\begin{equation}
\label{48}
\begin{array}{rcl}
\displaystyle
\frac{d\phi_{10}(t)}{d t}+\frac{d\psi_{0}(t)}{d t}\phi_{11,-}&= &
2u_0+e_1+e_2,\\
\displaystyle
\frac{d\phi_{20}(t)}{d t}+\frac{d\psi_{0}(t)}{d t}\phi_{21,-}&= &
2u_0+e_2+e_1.\\
\end{array}
\end{equation}
Remind that here the functions $\phi_{k0}(t)$ are determined by equations
(\ref{37}) for all $t\ge 0$.

    It is clear from (\ref{48}) that phase limit values of both shock waves
coincide, that is,
\begin{equation}
\label{49}
\widehat{\phi}_{2,-}(t)=\widehat{\phi}_{1,-}(t)
\stackrel{def}{=}\widehat{\phi}_{-}(t).
\end{equation}

    Thus, it follows from (\ref{48}), (\ref{49}) that after the interaction,
i.e. for $t>t^*$, the discontinuities merge together constituting a new
shock wave whose dynamics is determined by the equation
\begin{equation}
\label{50}
\frac{d\widehat{\phi}_{-}(t)}{dt}=2u_0+e_1+e_2
\end{equation}
with the initial value $\widehat{\phi}_{-}(t^*)= \phi_{k0}(t^*)$.

    Equation (\ref{50}) represents the Hugoniot condition (\ref{21}) along
the discontinuity line after the interaction, since
$[u]= u_0-(u_0 + e_1 + e_2)= -(e_1 + e_2)$ and
$[f(u)]= (u_0)^2-(u_0 + e_1 + e_2)^2$.

    3. Now we describe the dynamics of the shock wave merging process.
To this end, substituting the phases $\phi_{k0}(t)$ from (\ref{36})
into (\ref{43}), and using the relations $B_1(\rho)+B_2(-\rho)= 1$
from (\ref{27}) and $\psi_{0t}(t)= -\Big(e_1+e_2\Big)$, we derive the
equations determining the perturbations for the phases $\phi_{k1}(\tau)$:
$$
\displaystyle
\frac{d}{d\tau}\Big[\tau\phi_{11}(\tau)\Big]=
(-1)^{k-1}\frac{2e_{3-k}}{e_1+e_2}\Big(1-B_1(\rho)\Big), \quad k= 1,2.
$$

    Integrating these equations, we have the following expression for
the phase perturbations
\begin{equation}
\label{51}
\phi_{k1}(\tau)= (-1)^{k-1}\frac{2e_{3-k}}{(e_1+e_2)\tau}\int_{0}^{\tau}
\Big(1-B_1\big(\rho(\tau')\big)\Big)\,d\tau', \quad k= 1,2,
\end{equation}
where $\rho=\rho(\tau)$ solves the differential equation with the boundary
condition (\ref{44}), (\ref{45}):
$$
\begin{array}{rcl}
\displaystyle
\frac{d\rho}{d\tau} &=& 2B_1(\rho)-1, \\
\displaystyle
\frac{\rho(\tau)}{\tau}\Big|_{\tau \to +\infty}&=&1.\\
\end{array}
$$

    Verify that the functions $\phi_{k1}(\tau)$ found in (\ref{51})
satisfy the required properties.

    We have chosen the approximating functions $\omega_k(z)$ so that they
either have compact supports or decrease sufficiently rapidly
as $|z|\to \infty$, for example, $|\omega_k(z)|\le C_k(1+|z|)^{-N}$,
\ $N= 1,2,\dots$, \ $k= 1,2$. Therefore, we have
$\big({\check \omega}_1*\omega_2\big)(\eta) \le K(1+|z|)^{-N}$ as
$|z|\to \infty$, \ $N= 1,2,\dots$ and $B_1(\rho)=
\int_{-\infty}^{\rho}\big({\check \omega}_1*\omega_2\big)(\eta)\,d\eta
= 1+O(\rho^{-N})$ as $\rho \to +\infty$, \ $B_1(\rho)= O(\rho^{-N})$ as
$\rho \to -\infty$.

    If $\tau \to \infty$ then $\rho \to \infty$ and, according to
(\ref{51}) and Remark~\ref{rem22}, we have
$$
\begin{array}{rcl}
\displaystyle
\phi_{k1}(\tau) &= & O(\tau^{-1}),\\
\displaystyle
\tau\frac{d\phi_{k1}(\tau)}{d\tau} &= & O(\tau^{-1}).\\
\end{array}
$$

    Note that $\rho= \rho_0$ is an ordinary (nonmultiple) root of the
right-hand side of the differential equation (\ref{44}), (\ref{45}).
Indeed, $F(\rho_0)= 2B_1(\rho_0)-1= 0$ and accordingly to (\ref{26}),
$$
F'(\rho_0)= 2\big({\check \omega}_1*\omega_2\big)(\rho_0)
= 2\int_{-\infty}^{\infty}\omega_k(\xi)\omega_{3-k}(\rho_0+\xi)\,d\xi>0
$$
since $\omega_k(\xi)\ge 0$, \ $k= 1,2$.

    Thus, if $\tau \to -\infty$ then $\rho \to \rho_0$ and according to
Proposition~\ref{pr1}, $\rho-\rho_0= O(|\tau|^{-N})$ for all $N=1,2,\dots$.
Therefore, using Remark~\ref{rem22}, we have
$B_1(\rho)= 1/2+O(|\tau|^{-N})$, \ $\tau \to -\infty$.
According to (\ref{51}), we obtain
$$
\begin{array}{rcl}
\displaystyle
\phi_{k1}(\tau) &= &(-1)^{k-1}\frac{e_{3-k}}{e_1+e_2}+ O(\tau^{-1}),\\
\displaystyle
\psi_{1}(\tau) &= &-1+O(\tau^{-1}).\\
\end{array}
$$

    Calculate the limit of the expression
$\tau\frac{d\phi_{k1}(\tau)}{d\tau}$ as $\tau \to -\infty$. To this end,
rewrite this expression in the form
$$
\tau\frac{d\phi_{k1}(\tau)}{d\tau}
= (-1)^{k-1}\frac{e_{3-k}}{e_1+e_2}\big(1-B_1(\rho)\big)
\bigg[1-\frac{\int_{0}^{\tau}\Big(1-B_1\big(\rho(\tau')\big)\Big)\,d\tau'}
{\tau\big(1-B_1(\rho)\big)}\bigg].
$$
Applying the L'Hospital rule to the limit in square brackets we find
$$
\lim_{\tau \to -\infty}\tau\phi_{k1\tau}(\tau)= 0.
$$

    As one can see, the functions $\phi_{k1}(\tau)$ obtained in (\ref{51})
have all the properties required.

    Thus, we have proved the Theorem~\ref{th3}.

\section{Proof of Theorem~1.4.\/}
\label{s5}

    1. We consider the case when $u_0(x,t)$ and $e_k(x,t)$ in ansatz
(\ref{8}) are constants. In this case, therefore the interaction switch
functions from Lemma~\ref{lem2} $B_k(x,t,\rho)= B_k(\rho)$. As in
Section~\ref{s4}, we shall seek shock wave phases in ansatz (\ref{8})
as functions of $\varepsilon $, i.e. $\phi_k= \phi_k(t,\varepsilon )$,
\ $k= 1,2$. Denote the distance between shock wave fronts by
$(\psi(t,\varepsilon )=\phi_2(t,\varepsilon )-\phi_1(t,\varepsilon )$.

    Substituting the singular ansatz (\ref{8}) and asymptotics (\ref{65}),
given by Lemma~\ref{lem2}, into equation (\ref{1}) we have
$$
L[u^*(x,t,\varepsilon )]
=u^*_{{\varepsilon }t} + [f(u^*_{\varepsilon })]_x
\qquad\qquad\qquad\qquad\qquad\qquad\qquad\qquad\qquad\qquad\qquad
$$
$$
+\sum_{k= 1}^2\Big\{e_k\frac{d\phi_{k}(t,\varepsilon )}{dt}
-\Big[f(u_0+e_k)-f(u_0)\Big]
-B_k\big((-1)^{k-1}\rho\big)\Big\}\delta(-x+\phi_k(t,\varepsilon ))
$$
\begin{equation}
\label{75}
\qquad\qquad\qquad\qquad\qquad\qquad\qquad\qquad\qquad\qquad
+O_{{\cal D}'}(\varepsilon ), \quad \varepsilon  \to +0.
\end{equation}
where $\rho=\frac{\psi(t,\varepsilon )}{\varepsilon }$ and the estimate
$O_{{\cal D}'}(\varepsilon )$ in this representation is uniform with
respect to the distance $\psi(t,\varepsilon )$.

     From (\ref{75}), using Lemma~\ref{lem4}, we derive the necessary
and sufficient conditions for the relation
$L[u^*_{\varepsilon }(x,t)]= O_{{\cal D}'}(\varepsilon )$ to be valid:
\begin{equation}
\label{76}
\begin{array}{rcl}
\displaystyle
e_1\frac{d}{d t}\phi_{1}(t,\varepsilon )&=&
\Big[f(u_0+e_1)-f(u_0)\Big]+B_1\big(\rho\big), \\
\displaystyle
e_2\frac{d}{d t}\phi_{2}(t,\varepsilon )&=&
\Big[f(u_0+e_2)-f(u_0)\Big]+B_2\big(-\rho\big). \\
\end{array}
\end{equation}

    When
$$
\psi(t,\varepsilon )= \phi_2(t,\varepsilon )-\phi_1(t,\varepsilon )
> c{\varepsilon }^{1-\alpha},
$$
where $c>0$, \ $0< \alpha \le 1$, that is, before the interaction
of shock waves, according to Remark~\ref{rem22} and (\ref{67}), we
have, up to $O(\varepsilon )$:
$$
\begin{array}{rcl}
\displaystyle
B_1\big(\frac{\psi}{\varepsilon }\big)&=&
f(u_0+e_1+e_2)-f(u_0+e_1)-f(u_0+e_2)+f(u_0), \\
\displaystyle
B_2\big(-\frac{\psi}{\varepsilon }\big)&=&0. \\
\end{array}
$$

    It follows from the above that system (\ref{76}) turns into the
system of equations which, according to Theorem~\ref{th2}, describes
the dynamics of two noninteracting shock waves:

\begin{equation}
\label{77}
\begin{array}{rcl}
\displaystyle
e_1\frac{d\phi_{10}(t)}{dt}&= &f(u_0+e_1+e_2)-f(u_0+e_2), \\
\displaystyle
e_2\frac{d\phi_{20}(t)}{dt}&= &f(u_0+e_2)-f(u_0). \\
\end{array}
\end{equation}

    Equations (\ref{77}) obtained above represent the Hugoniot conditions
(\ref{21}) along the lines: $[u]=e_1$, \ $[f(u)]=f(u_0+e_1+e_2)-f(u_0+e_2)$
for the lagging shock wave, and $[u]=e_2$, \ $[f(u)]=f(u_0+e_2)-f(u_0)$
for the advanced one.

    Denote by $\psi_0(t)= \phi_{20}(t)-\phi_{10}(t)$ the distance between
the shock waves before the instant of the interaction $t=t^*$ which we
define as a solution of the equation $\psi_0(t^*)=0$.

    Thus, before the instant of the interaction, as $t\in (0, \ t^*)$,
the shock waves move with velocities (\ref{77}) along the straight lines

\begin{equation}
\label{78}
\begin{array}{rcl}
\displaystyle
x_1= \phi_{10}(t)&= &x_1^0+\frac{f(u_0+e_1+e_2)-f(u_0+e_2)}{e_1}t, \\
\displaystyle
x_2= \phi_{20}(t)&= &x_2^0+\frac{f(u_0+e_2)-f(u_0)}{e_2}t, \\
\end{array}
\end{equation}
which intersect at a point with the following coordinates:
\begin{equation}
\label{78*}
\begin{array}{rcl}
\displaystyle
t^* &= &
e_1e_2\frac{x_2^0-x_1^0}{e_2f(u_0+e_1+e_2)-(e_1+e_2)f(u_0+e_2)+e_1f(u_0)},\\
\displaystyle
x^*&= &\frac{\Big[f(u_0+e_1+e_2)-f(u_0+e_2)\Big]e_2x_2^0
-\Big[f(u_0+e_2)-f(u_0)\Big]e_1x_1^0}
{e_2f(u_0+e_1+e_2)-(e_1+e_2)f(u_0+e_2)+e_1f(u_0)},\\
\end{array}
\end{equation}
where $\phi_{k0}(0)$ are the initial positions of the discontinuities.

    2. As in Section~\ref{s4}, define shock wave phases by formula
(\ref{38}) where the functions $\phi_{k0}(t)$ determined by equations
(\ref{78}), as $t\in (0, \ t^*)$, {\it are expanded by the same
equations\/} (\ref{78}) for $t\in (t^*, \ +\infty)$, as was done
previously, and the phase derivatives with respect to time are given
by formulae (\ref{41}). The full phase difference is
$$
\psi(t,\varepsilon )= \psi_{0}(t)\big(1 + \psi_{1}(\tau)\big),
$$
where $\psi_{1}(\tau)= \phi_{21}(\tau)-\phi_{11}(\tau)$. Here the
independent variable of the interaction switch function is set in (\ref{40}).

    For $\phi_{k1}(\tau)$ we set the boundary conditions (\ref{39}).
The limit values of phases and their derivatives with respect to time, as
$\tau \to -\infty$, were determined in (\ref{42}).

    Substituting (\ref{38}), (\ref{41}) into (\ref{76}), we obtain, for
all $t> 0$ and $\tau \in (-\infty, \ \infty)$, the system of equations with
boundary conditions (\ref{39}):
\begin{equation}
\label{79}
\begin{array}{rcl}
\displaystyle
e_1\Big(\frac{d\phi_{10}(t)}{d t}
+\frac{d\psi_{0}(t)}{dt}\frac{d}{d\tau}\Big[\tau\phi_{11}(\tau)\Big]\Big)
&=& f(u_0+e_1)-f(u_0) + B_1\big(\rho\big), \\
\displaystyle
e_2\Big(\frac{d\phi_{20}(t)}{d t}
+\frac{d\psi_{0}(t)}{dt}\frac{d}{d\tau}\Big[\tau\phi_{21}(\tau)\Big]\Big)
&=& f(u_0+e_2)-f(u_0) + B_2\big(-\rho\big), \\
\end{array}
\end{equation}
where $\phi_{k0}(t)$ are the extensions of the functions $\phi_{k0}(t)$
determined by equations (\ref{77}) for all $t\ge t^*$, the extension
being determined by the same equations (\ref{77}).

    Subtracting one of the equations in system (\ref{79}) from the other
we obtain the following differential equation with the boundary condition
for $\rho(\tau)$:
\begin{equation}
\label{80}
\begin{array}{rcl}
\displaystyle
\rho_{\tau}&= &F(\rho), \\
\displaystyle
\frac{\rho}{\tau}\Big|_{\tau \to +\infty}&= &1, \\
\end{array}
\end{equation}
where
\begin{equation}
\label{81}
\displaystyle
F(\rho)= \frac{1}{\frac{d\psi_{0}(t)}{d t}}
\bigg[\frac{f(u_0+e_2)-f(u_0)+B_2(-\rho)}{e_2}
-\frac{f(u_0+e_1)-f(u_0)+B_1(\rho)}{e_1}\bigg],
\end{equation}
here the boundary condition follows from (\ref{39}).

    Formula (\ref{77}) implies:
\begin{equation}
\label{82}
\psi_{0t}(t)
= \frac{f(u_0+e_2)-f(u_0)}{e_2}-\frac{f(u_0+e_1+e_2)-f(u_0+e_2)}{e_1}.
\end{equation}

    Taking into account the first correlation from (\ref{67}), convert
function (\ref{81}) to the form
$$
F(\rho)
\qquad\qquad\qquad\qquad\qquad\qquad\qquad\qquad\qquad\qquad\qquad\qquad
\qquad\qquad\qquad\qquad\qquad
$$
\begin{equation}
\label{83}
=\frac{\Big[f(u_0+e_1+e_2)-f(u_0+e_1)\Big]e_1-\Big[f(u_0+e_1)-f(u_0)\Big]e_2
-(e_1+e_2)B_1(\rho)}
{\Big[f(u_0+e_2)-f(u_0)\Big]e_1-\Big[f(u_0+e_1+e_2)-f(u_0+e_2)\Big]e_2}.
\end{equation}

    Using the limit values of the switch function $B_1(\pm\infty)$ from
Lemma~\ref{lem2} we find the limit values of the function $F(\rho)$ from
(\ref{83}):
\begin{equation}
\label{84}
\begin{array}{rcl}
\displaystyle
F(\rho)\Big|_{\rho \to +\infty}&=&1, \\
\displaystyle
F(\rho)\Big|_{\rho \to -\infty}&=&
\frac{\Big[f(u_0+e_1+e_2)-f(u_0+e_1)\Big]e_1-\Big[f(u_0+e_1)-f(u_0)\Big]e_2}
{\Big[f(u_0+e_2)-f(u_0)\Big]e_1-\Big[f(u_0+e_1+e_2)-f(u_0+e_2)\Big]e_2}.
\end{array}
\end{equation}

    Rewrite the limit value $F(\rho)\Big|_{\rho \to -\infty}$ from
(\ref{84}) in the form
$$
F(\rho)\Big|_{\rho \to -\infty}
= -\frac{f(u_0+e_1+e_2)e_1-f(u_0+e_1)(e_1+e_2)+f(u_0)e_2}
{f(u_0+e_1+e_2)e_2-f(u_0+e_2)(e_1+e_2)+f(u_0)e_1}.
$$

    In view of our assumptions, the function $f(u)$ is convex downwards,
that is, $f^{\prime\prime}(u) > 0$ on the range of the solution~$u$,
and the amplitudes are positive,
$e_1, \ e_2>0$. Under these assumptions, both the numerator and denominator
of the last fraction are strictly positive, since for a function convex
downwards the inequality $(x_2-x)f(x_1)+(x_1-x_2)f(x)+(x-x_1)f(x_2)> 0$
holds for $x_1 < x < x_2$.
In our case, $x_1= u_0$, \ $x= u_0+e_1$, \
$x_2= u_0+e_1+e_2$.

    Thus, as $\rho \to \pm\infty$, the limit values of the right hand
side $F(\rho)$ of the differential equation (\ref{80}) have opposite signs.

    According to (\ref{66}), the derivative of the interaction switch
function has the form
$$
\frac{d}{d\rho}B_1(\rho)
= e_1e_2\int f^{\prime\prime}\big(u_0+e_1\omega_{01}(-\eta)
+e_2\omega_{02}(-\eta+\rho)\big)
\omega_{1}(-\eta)\omega_{2}(-\eta+\rho)\,d\eta.
$$
It is positive since $f^{\prime\prime}(u) \ge 0$, \ $\omega_{k}(\eta)>0$
and $e_1, \ e_2>0$. It follows from here and from (\ref{83}) that the
derivative
\begin{equation}
\label{84*}
\frac{d}{d\rho}F(\rho)
= \frac{(e_1+e_2)}{f(u_0+e_1+e_2)e_2-f(u_0+e_2)(e_1+e_2)+f(u_0)e_1}
\frac{d}{d\rho}B_1(\rho)
\end{equation}
is positive, and therefore $F(\rho)$ is an increasing function.

    Since the function $F(\rho)$ is smooth, the equation $F(\rho)= 0$ has
a root $\rho_0$. Then, according to Proposition~\ref{pr1}, the solution
$\rho$ of equation (\ref{80}) tends to $\rho_0$, as $\tau \to -\infty$.

   Due to the smoothness of the function $F(\rho)$ and the aforesaid,
the equation $F(\rho)= 0$ has at least one root $\rho_0$. Let $\rho_0$ be
the maximal root of the equation $F(\rho)= 0$. Then, according to
Proposition~\ref{pr1}, as $\tau \to -\infty$, the solution
of equation (\ref{80}), $\rho$ tends to $\rho_0$.

    Pass to the limit in (\ref{79}), as $\tau \to -\infty$, taking into
account that by (\ref{39}), $\tau\phi_{k1\tau}(\tau) \to 0$. From here we
derive a system of equation describing the evolution of shock waves after
the interaction, for $t>t^*$:
\begin{equation}
\label{85}
\begin{array}{rcl}
\displaystyle
\frac{d\phi_{10}(t)}{d t}+\phi_{11,-}\frac{d\psi_{0}(t)}{d t} &= &
\frac{f(u_0+e_1)-f(u_0) + B_1\big(\rho_0\big)}{e_1}, \\
\displaystyle
\frac{d\phi_{20}(t)}{d t}+\phi_{21,-}\frac{d\psi_{0}(t)}{d t} &= &
\frac{f(u_0+e_2)-f(u_0) + B_2\big(-\rho_0\big)}{e_2}. \\
\end{array}
\end{equation}

    As shown above, we have
$\rho= \tau\big(1+\psi_{1}(\tau)\big) \to \rho_0$ as $\tau \to -\infty$.
It follows that $\psi_{1}(\tau) \to -1$, that is
$\psi_{1}(\tau)= -1+O(\tau^{-1})$.

    Since $\psi_0(t)= \phi_{20}(t)-\phi_{10}(t)$ and
$\psi_0(t)O(\tau^{-1})= O(\varepsilon )$, we have as $\tau \to -\infty$:
$$
\widehat{\phi}_{2}(\tau)= \phi_{20}(t)+\psi_0(t)\phi_{21}(\tau)
$$
$$
= \phi_{20}(t)+\psi_0(t)\bigg(\phi_{11}(\tau) -1 + O(\tau^{-1})\bigg)
= \widehat{\phi}_{1}(\tau)+O(\varepsilon ).
$$
That is, up to $O(\varepsilon )$ we have
\begin{equation}
\label{86}
\widehat{\phi}_{2,-}(t)= \widehat{\phi}_{1,-}(t)
\stackrel{def}{=}\widehat{\phi}_{-}(t).
\end{equation}

     It is clear from (\ref{85}), (\ref{86}) that after the interaction,
i.e. for $t>t^*$, discontinuities merge constituting a new shock wave
whose dynamics is determined by the equation
\begin{equation}
\label{87}
\frac{d\widehat{\phi}_{-}(t)}{dt}
=\frac{f(u_0+e_1)-f(u_0) + B_1\big(\rho_0\big)}{e_1}.
\end{equation}

    Here the following relation holds
\begin{equation}
\label{88}
\frac{\Big[f(u_0+e_1)-f(u_0)\Big] + B_1\big(\rho_0\big)}{e_1}
= \frac{\Big[f(u_0+e_2)-f(u_0)\Big] + B_2\big(-\rho_0\big)}{e_2}.
\end{equation}

    From (\ref{88}) using the first relation from (\ref{67}), by
Lemma~\ref{lem2}, we find the value
\begin{equation}
\label{89}
B_1(\rho_0)= \frac{\Big[f\big(u_0+e_1+e_2\big)-f\big(u_0+e_1\big)\Big]e_1
-\Big[f\big(u_0+e_1\big)-f\big(u_0\big)\Big]e_2}{e_1+e_2}.
\end{equation}

     Substituting value (\ref{89}) into (\ref{87}) we obtain the equation
which describes the propagation of the shock wave resulting from merging
the initial waves:
\begin{equation}
\label{90}
\frac{d\widehat{\phi}_{-}(t)}{dt}=
\frac{f\big(u_0+e_1+e_2\big)-f\big(u_0\big)}{e_1+e_2}
\end{equation}
with the initial value $\widehat{\phi}_{-}(t^*)= \phi_{k0}(t^*)= x^*$.

    Equation (\ref{90}) is the Hugoniot condition (\ref{21}) along
the line of discontinuity after merging the shock waves, since
$[u]= u_0-(u_0 + e_1 + e_2)= -(e_1 + e_2)$ and
$[f(u)]= f\big(u_0+e_1+e_2\big)-f\big(u_0\big)$.

    3. Consider the dynamics of the process of merging two shock waves
into a new one. Substitute the phases of noninteracting shock waves
$\phi_{k0}(t)$ from (\ref{77}) into (\ref{79}). From here, using
the first correlation from (\ref{67}) for the interaction switch
functions $B_k(\rho)$ and expression (\ref{77}) for $\psi_{0t}(t)$ we
obtain equations to determine $\phi_{k1}(\tau)$, \ $k= 1,2$:
$$
\frac{d}{d\tau}\Big[\tau\phi_{k1}(\tau)\Big]
\qquad\qquad\qquad\qquad\qquad\qquad\qquad\qquad\qquad\qquad\qquad\qquad
\qquad\qquad\qquad\qquad
$$
\begin{equation}
\label{91}
= (-1)^{k-1}e_{3-k}
\frac{f\big(u_0+e_1+e_2\big)-f\big(u_0+e_1\big)-f\big(u_0+e_2\big)
+f\big(u_0\big)-B_1(\rho(\tau))}
{\Big[f\big(u_0+e_1+e_2\big)-f\big(u_0+e_2\big)\Big]e_2
-\Big[f\big(u_0+e_2\big)-f\big(u_0\big)\Big]e_1}.
\end{equation}
Integrating equations (\ref{91}) we derive the following expression for
$\phi_{k1}(\tau)$:
$$
\phi_{k1}(\tau)
\qquad\qquad\qquad\qquad\qquad\qquad\qquad\qquad\qquad\qquad\qquad\qquad
\qquad\qquad\qquad\qquad\qquad\qquad\qquad
$$
\begin{equation}
\label{92}
= (-1)^{k-1}\frac{e_{3-k}}{\tau}\int\limits_{0}^{\tau}
\frac{f\big(u_0+e_1+e_2\big)-f\big(u_0+e_1\big)-f\big(u_0+e_2\big)
+f\big(u_0\big)-B_1(\rho(\tau'))}
{\Big[f\big(u_0+e_1+e_2\big)-f\big(u_0+e_2\big)\Big]e_2
-\Big[f\big(u_0+e_2\big)-f\big(u_0\big)\Big]e_1}\,d\tau',
\end{equation}
where $\rho= \rho(\tau)$ solves the differential equation with boundary
condition (\ref{80}), (\ref{81}).

    As in the case of the Hopf equation, it can be verified that the
functions $\phi_{k1}(\tau)$ found in (\ref{92}) satisfy the presupposed
properties.

    Let $\tau \to \infty$, then $\rho \to \infty$, \ $\rho \sim \tau$
and, according to (\ref{92}) and Remark~\ref{rem22},
$$
\begin{array}{rcl}
\displaystyle
\phi_{k1}(\tau) &= & O(\tau^{-1}),\\
\displaystyle
\tau\frac{d\phi_{k1}(\tau)}{d\tau} &= & O(\tau^{-1}).\\
\end{array}
$$

   According to (\ref{89}), $\rho= \rho_0$ is an ordinary (nonmultiple)
root of the right-hand side of the differential equation (\ref{80}),
(\ref{83}), i.e. $F(\rho_0)= 0$. But in view of to (\ref{84*}),
$\frac{d}{d\rho}F(\rho)>0$. Consequently, $\rho= \rho_0$ is an ordinary
(nonmultiple) root of the right-hand side of the differential equation
(\ref{80}), (\ref{83}).

    Thus, if $\tau \to -\infty$, then $\rho \to \rho_0$ and $B_1(\rho)$
tends to the limit value $B_1(\rho_0)$ given by formula (\ref{89}).
According to Proposition~\ref{pr1} $\rho-\rho_0= O(|\tau|^{-N})$ for
all $N= 1,2,\dots$. Thus, using Remark~\ref{rem22}, we have
$B_1(\rho)= B_1(\rho_0)+O(|\tau|^{-N})$, \ $\tau \to -\infty$.
Therefore, (\ref{91}), (\ref{92}) imply the same estimates as in the
case of the Hopf equation:
$$
\begin{array}{rcl}
\displaystyle
\phi_{k1}(\tau) &=&(-1)^{k-1}\frac{e_{3-k}}{e_1+e_2}+O(\tau^{-1}),\\
\displaystyle
\psi_{1}(\tau) &=&-1+O(\tau^{-1}),\\
\displaystyle
\tau\frac{d\phi_{k1}(\tau)}{d\tau}&=&O(\tau^{-1}). \\
\end{array}
$$

    Thus, the functions $\phi_{k1}(\tau)$ constructed in (\ref{92}) have
all presupposed properties.

    Thus, we have proved the Theorem~\ref{th4}.

    The authors are greatly indebted to E.~Yu.~Panov for many fruitful
discussions.


\begin{thebibliography}{15}


\bibitem{1} O.~A.~Oleinik
\newblock Discontinuous solutions of nonlinear equations.
\newblock Usp. Math. Nauk. (N.S.), 1957, v.12, N 3, 3--73;
English transl. in Amer. Math. Soc. Transl. Ser., 2, {\bf 26}, 95--172.

\bibitem{2} I.~M.~Gelfand
\newblock Some problems in the theory of quasilinear equations.
\newblock Usp. Math. Nauk. (N.S.), 1959, v.14, N 2, 87--158;
English transl. in Amer. Math. Soc. Transl. Ser., 2, (1963), 295--381.

\bibitem{3} Joel Smoller
\newblock Shock Waves and Reaction-Diffusion Equations.
\newblock Springer-Verlag, 1983.

\bibitem{4} B.~L.~Rozdestvenskii and N.~N.~Yanenko
\newblock Systems of Quasi-Linear Equations.
\newblock Nauka, Moscow, 1978.

\bibitem{5} G.~B.~Whitham
\newblock Linear and Nonlinear waves.
\newblock John Wiley and Sons, New York, London, Toronto, 1974.

\bibitem{6} V.~P.~Maslov
\newblock Propagation of shock waves in isoentropic gas
\newblock Contemporary Problems in Mathematics.
Fundamental Directions. Itogi Nauki i Tekhniki, v. 8,
Moscow, Publ VINITI AN SSSR, 1977, 199--272.

\bibitem{7} V.~P.~Maslov and V.~A.~Tsupin
\newblock Nessesary conditions for the existence of infinetely
narrow solitons in gas dynamics
\newblock (Russian) Dokl. Akad. Nauk SSSR, 1979, v. 246, 298--300;
Soviet Phys. Dokl., 1979, v.24, N 5, 354--356.

\bibitem{8} V.~P.~Maslov and G.~A.~Omel'yanov
\newblock Asymptotic soliton-form solutions of equations with
small dispersion.
\newblock Russian Math. Surveys, 1981, v. 36, N 3, 1981, 73-119;
translated from Uspekhi Mat. Nauk., 1981, v. 36, N 3, 63--126.

\bibitem{9} V.~G.~Danilov, V.~P.~Maslov, V.~M.~Shelkovich.
\newblock Algebra of singularities of singular solutions to first-order
quasilinear strictly hyperbolic systems,
\newblock Theor. Math. Phys., 1998, v. 114, N 1, 1--42.

\bibitem{10} J.~Rauch and M.~Reed
\newblock Nonlinear superposition and absorption of delta waves in
in one Space dimension,
\newblock J. Funct. Anal., 1987, v.73, 152--178.

\bibitem{11} M.~Oberguggenberger
\newblock Multiplication of distributions and applications to partial
differential equations.
\newblock N.-Y.,1992.

\bibitem{12} T.~Gramchev
\newblock Semilinear Hyperbolic Systems and Equations with Singular
Initial Data.
\newblock Mh. Math., 1991, v.112, 99--113.

\bibitem{13} J.~F.~Colombeau
\newblock Elementary introduction to new generalized functions.
\newblock North Holland, 1985.

\bibitem{14} Yu.~V.~Egorov
\newblock A contribution to the theory of generalized functions
\newblock Russian Math. Surveys, 1990, v. 45, N 5, 1981, 1-49;
translated from Uspekhi Mat. Nauk., 1990, v.45, N 5, 3--40.

\bibitem{15} Ya.~B.~Livchak
\newblock Towards the theory of generalized functions.
\newblock Trudy Rizhskogo Algebr. Seminar, Riga 1969, 98--164 (in Russian).

\bibitem{16} Li Bang-He.
\newblock Non-standard analysis and multiplication of distributions
\newblock Acta scientia sinica, 1978, v.21, N 5, 561--585.

\bibitem{17} V.~K.~Ivanov
\newblock Asymptotical approximation to the product of generalized functions.
\newblock Izv. Vyssh. Uchebn. Zaved. Mat., 1981, N 1, 19--26 (in Russian).

\bibitem{18} V.~M.~Shelkovich.
\newblock  An associative algebra of distributions and multipliers,
\newblock (Russian) Dokl. Akad. Nauk SSSR, 1990, v. 314, 159--164;
translated from Soviet Math. Dokl., 1991, v. 42, 409--414.

\bibitem{19} V.~M.~Shelkovich.
\newblock An associative-commutative algebra of distributions that
includes multiplicators, generalized solutions of nonlinear equations,
\newblock Mathematical Notices, 1995, v. 57, N 5, 765--783.

\bibitem{20} V.~S.~Vladimirov
\newblock Generalized Functions in Mathematical Physics.
\newblock Mir Publ., Moscow, 1979.

\end{thebibliography}
\end{document}